\documentclass[lettersize,journal]{IEEEtran}

\usepackage{amsmath,amsfonts}
\usepackage{algorithmic}
\usepackage{algorithm}
\usepackage{array}
\usepackage{textcomp}
\usepackage{stfloats}
\usepackage{url}
\usepackage{verbatim}
\usepackage{graphicx}
\usepackage{cite}
\usepackage{makecell}
\usepackage{rotating}
\usepackage[dvipsnames,table]{xcolor}
\usepackage{soul}
\usepackage{balance}
\usepackage[utf8]{inputenc}
\usepackage[T1]{fontenc}
\usepackage{hhline}
\usepackage{fixltx2e}
\usepackage{tabularx}
\usepackage{booktabs}
\usepackage{multirow}
\usepackage{enumitem}
\usepackage{flushend}
\usepackage{pifont}
\usepackage[most]{tcolorbox}
\usepackage[scaled=0.86]{helvet}
\usepackage{xspace}
\usepackage{listings}

\usepackage[colorlinks,bookmarksopen,bookmarksnumbered,citecolor=red, urlcolor=black]{hyperref}
\usepackage{cite}
\usepackage{lscape}
\usepackage[framemethod=TikZ]{mdframed}
\usepackage{tabularray}

\definecolor{light-gray}{gray}{0.9}
\mdfdefinestyle{MyFrame}{
    linecolor=black,
    outerlinewidth=0.15pt,
    roundcorner=3pt,
    innertopmargin=2pt,
    innerbottommargin=2pt,
    innerrightmargin=4pt,
    innerleftmargin=4pt,
    backgroundcolor=light-gray}
    
\mdfdefinestyle{MyFrameSimple}{
    linecolor=black,
    outerlinewidth=0.15pt,
    roundcorner=3pt,
    innertopmargin=2pt,
    innerbottommargin=2pt,
    innerrightmargin=4pt,
    innerleftmargin=4pt}

\newcommand{\sectopic}[1]{\vspace*{0.1em}\par\noindent{\textit{\bfseries #1}}}

\newcommand{\figref}[1]{Fig.~\ref{#1}}

\newcommand{\citep}[1]{\cite{#1}}
\newcommand{\addcitet}[2]{\csdef{mapcitet#1}{#2}}
\newcommand{\citet}[1]{\csuse{mapcitet#1}\cite{#1}}
\addcitet{storey2010impact}{Storey et~al.}
\addcitet{vasilescu2015gender}{Vasilescu et~al.}
\addcitet{Tsay2014}{Tsay et~al.}
\addcitet{ehsan2020empirical}{Ehsan et~al.}
\addcitet{arora2022ask}{Arora et~al.}
\addcitet{liu2018searching}{Liu et~al.}
\addcitet{beyer2018automatically}{Beyer et~al.}
\addcitet{kenton2019bert}{Kenton and Toutanova}
\addcitet{lill2024helpfulness}{Lill et~al.}
\addcitet{li_ner_2022}{Li et~al.}
\addcitet{ye2016software}{Ye et~al.}
\addcitet{huang_intent_2020}{Huang et~al.}
\addcitet{subash2022disco}{Subash et~al.}
\addcitet{jiang2024mixtral}{Jiang et~al.}
\addcitet{mixtral2024}{AI}
\addcitet{pouya2024replication}{Fathollahzadeh et~al.}
\addcitet{artificialanalysis}{{Artificial Analysis}}
\addcitet{OpenAiFinetuning}{{OpenAi}()}
\addcitet{Chatcompletion}{{Huggingface}()}
\addcitet{tabassum2020code}{Tabassum et~al.}
\addcitet{das2023zero}{Das et~al.}
\addcitet{veera2019nerse}{Veera Prathap~Reddy et~al.}
\addcitet{rosen2016mobile}{Rosen and Shihab}
\addcitet{allamanis2013and}{Allamanis and Sutton}
\addcitet{treude2011programmers}{Treude et~al.}
\addcitet{beyer2014manual}{Beyer and Pinzger}
\addcitet{beyer2017analyzing}{Beyer et~al.}
\addcitet{anderson2012discovering}{Anderson et~al.}
\addcitet{coleman1975computer}{Coleman and Liau}
\addcitet{yang2016security}{Yang et~al.}
\addcitet{mondal2021early}{Mondal et~al.}
\addcitet{vasilescu2014social}{Vasilescu et~al.}
\addcitet{backstrom2006group}{Backstrom et~al.}
\addcitet{calefato2018ask}{Calefato et~al.}
\addcitet{hutto2014vader}{Hutto and Gilbert}
\addcitet{guzman2013towards}{Guzman and Bruegge}
\addcitet{srba2016stack}{Srba and Bielikova}
\addcitet{cohen2013statistical}{Cohen}
\addcitet{hand2012assessing}{Hand}
\addcitet{mchugh2012interrater}{McHugh}
\addcitet{nguyen2010studying}{Nguyen et~al.}
\addcitet{noei2019too}{Noei et~al.}
\addcitet{noei2024detecting}{Noei et~al.}
\addcitet{pintas2021feature}{Pintas et~al.}
\addcitet{jiarpakdee2016study}{Jiarpakdee et~al.}
\addcitet{zhang2016variable}{Zhang}
\addcitet{mchugh2013chi}{McHugh}
\addcitet{lewis2020rag}{Lewis et~al.}
\addcitet{han2025graphrag}{Han et~al.}
\addcitet{hou2024large}{Hou et~al.}
\addcitet{fan2023large}{Fan et~al.}
\addcitet{zhang2020chatbot4qr}{Zhang et~al.}
\addcitet{preda2024supporting}{Preda et~al.}
\addcitet{hassani2025empirical}{Hassani et~al.}
\addcitet{guo2024exploring}{Guo et~al.}
\addcitet{colavito2024leveraging}{Colavito et~al.}
\addcitet{el2021exploring}{El~Mezouar et~al.}
\addcitet{shi2021first}{Shi et~al.}
\addcitet{tufano2024unveiling}{Tufano et~al.}

\newcommand{\dquote}[1]{``#1''}
\newcommand{\squote}[1]{`#1'}
\newcommand{\etal}{et~al.~}
\newcommand{\entity}[1]{\texttt{#1}}
\newcommand{\intent}[1]{\textit{#1}}
\newcommand{\ourmethod}{SENIR\xspace}

\newcommand{\rqone}{How effective is \ourmethod in labeling developer chatroom conversations?}
\newcommand{\rqtwo}{What features of the developer questions contribute to their resolution outcomes?}
\newcommand{\rqthree}{How do entities, intents, and their interactions impact the resolution of developer questions?}

\definecolor{custom-gray}{cmyk}{0,0,0,0.7,1.00}
\newtcolorbox{Summary}[2][]{
    top=0.15in,
    fonttitle=\bfseries,
    colbacktitle=custom-gray,
    colback=gray!5,
    colframe=gray!40!black,
    enhanced,
    attach boxed title to top left={xshift=1.5em,yshift=-\tcboxedtitleheight/2},
    boxed title style={size=small,colback=custom-gray},
    drop shadow={black!50!white},
    title=#2,#1}

\lstdefinelanguage{json}{
    basicstyle=\ttfamily\footnotesize,
    numbers=left,
    numberstyle=\scriptsize\color{gray},
    stepnumber=1,
    numbersep=8pt,
    showstringspaces=false,
    breaklines=true,
    frame=lines,
    backgroundcolor=\color{lightgray!10},
    literate=
     *{0}{{{\color{blue}0}}}{1}
      {1}{{{\color{blue}1}}}{1}
      {2}{{{\color{blue}2}}}{1}
      {3}{{{\color{blue}3}}}{1}
      {4}{{{\color{blue}4}}}{1}
      {5}{{{\color{blue}5}}}{1}
      {6}{{{\color{blue}6}}}{1}
      {7}{{{\color{blue}7}}}{1}
      {8}{{{\color{blue}8}}}{1}
      {9}{{{\color{blue}9}}}{1}
      {:}{{{\color{red}:}}}{1}
      {,}{{{\color{red},}}}{1}
      {\{}{{{\color{red}\{}}}{1}
      {\}}{{{\color{red}\}}}}{1}
      {[}{{{\color{red}[}}}{1}
      {]}{{{\color{red}]}}}{1},
}

\begin{document}

\title{Towards Refining Developer Questions using LLM-Based Named Entity Recognition for Developer Chatroom Conversations}

\author{
    Pouya Fathollahzadeh,~\IEEEmembership{Student Member,~IEEE,}
    Mariam El Mezouar,
    Hao Li, \\
    Ying Zou, and Ahmed E. Hassan,~\IEEEmembership{Fellow,~IEEE}
    \thanks{Pouya Fathollahzadeh and Ying Zou are with the Department of Electrical and Computer Engineering, Queen’s University, Kingston, ON K7L 3N6, Canada. E-mail: \{pouya.fathollahzadeh, ying.zou\}@queensu.ca.}
    \thanks{Hao Li and Ahmed E. Hassan are with the Software Analysis and Intelligence Lab (SAIL), School of Computing, Queen’s University, Kingston, ON K7L 3N6, Canada. E-mail: hao.li@queensu.ca, ahmed@cs.queensu.ca.}
    \thanks{Mariam El Mezouar is with the Department of Mathematics and Computer Science, Royal Military College of Canada, Kingston, ON K7K 7B4, Canada. E-mail: mariam.el-mezouar@rmc.ca.}
    % \thanks{Manuscript received [Your Date]; revised [Your Date].}
}

% Running head
\markboth{IEEE Transactions on Software Engineering, Vol.~XX, No.~XX, Month Year}
{Fathollahzadeh \MakeLowercase{\textit{et al.}}: Refining Developer Questions using LLM-Based NER}

\maketitle

\begin{abstract}
    %%%%%%%%%%%%%%%%%%%%%%%%%%%%%%%%%%%%%%%%%%%%%%
In software engineering chatrooms, communication is often hindered by imprecise questions that cannot be answered. Recognizing key entities~(e.g., programming languages and libraries) and user intent~(e.g., learning or requesting a review) can be essential for improving question clarity and facilitating better exchange. However, existing research using natural language processing techniques often overlooks these software-specific nuances.
In this paper, we introduce \underline{S}oftwar\underline{E}-specific \underline{N}amed entity recognition, \underline{I}ntent detection, and \underline{R}esolution classification~(SENIR), a labelling approach that leverages a Large Language Model to annotate entities, intents, and resolution status in developer chatroom conversations. To offer quantitative guidance for improving question clarity and resolvability, we build a resolution prediction model that leverages SENIR’s entity and intent labels along with additional predictive features.
We evaluate SENIR on the DISCO dataset using a subset of annotated chatroom dialogues. SENIR achieves an 86\% F-score for entity recognition, a 71\% F-score for intent detection, and an 89\% F-score for resolution status classification.
Furthermore, our resolution prediction model, tested with various sampling strategies~(random undersampling and oversampling with SMOTE) and evaluation methods~(5-fold cross-validation, 10-fold cross-validation, and bootstrapping), demonstrates AUC values ranging from 0.7 to 0.8. Key factors influencing resolution include positive sentiment and entities such as \entity{Programming Language} and \entity{User Variable} across multiple intents, while diagnostic entities~(e.g., \entity{Error Name}) are more relevant in error-related questions. Moreover, resolution rates vary significantly by intent: questions about \intent{API Usage} and \intent{API Change} achieve higher resolution rates, whereas \intent{Discrepancy} and \intent{Review} have lower resolution rates. A Chi-Square analysis confirms the statistical significance of these differences.

\end{abstract}


% Section files
\section{Introduction}\label{sec:introduction}

Developer chatrooms, such as Discord\footnote{\url{https://discord.com}} and Slack,\footnote{\url{https://slack.com}} are crucial tools for collaboration and knowledge sharing in software development. These platforms enable developers to seek help, solve technical problems, and engage continuously with the community. Collective knowledge available in chatrooms has been shown to accelerate software development and improve project outcomes~\citep{storey2010impact, vasilescu2015gender}. Additionally, chatrooms help build and maintain active developer communities, which are vital to the long-term success of open-source projects~\cite{Tsay2014}.

Despite the important role of chatrooms, their effectiveness is often hindered by communication issues. Poorly articulated questions that lack clarity or essential details frequently lead to misunderstandings, incomplete responses, or no response at all. For instance, research on Gitter\footnote{\url{https://gitter.im}} chatrooms finds that around 40\% of questions remain unanswered~\cite{ehsan2020empirical} and delays in resolving problems discourage community participation~\cite{arora2022ask}. While question-and-answer~(Q\&A) platforms like Stack Overflow are well studied~\citep{liu2018searching, beyer2018automatically, kenton2019bert}, research on developer chatrooms is still limited. Initial findings suggest that including URLs can reduce response rates, whereas user mentions improve the rates~\cite{ehsan2020empirical}. Recent work by~\citet{lill2024helpfulness} has explored automated methods to improve developer support in chatrooms by identifying similar past conversations. 

In developer chatrooms, a \textit{conversation} consists of an initial question or statement (e.g., asking for help or seeking clarification) followed by responses, clarifications, and follow-ups until the question is resolved or left unresolved. The quality of the initial question impacts resolution outcomes, as well-formed questions often include sufficient technical details, such as code snippets or error types, making them easier to address. Named Entity Recognition~(NER) is a useful method for extracting such technical details by identifying and categorizing domain-specific entities, such as libraries and error messages~\citep{li_ner_2022, ye2016software}. While traditional NER methods work well on structured and formal text, they often require extensive feature engineering and task-specific training datasets, limiting their adaptability to informal and fragmented chatroom conversations. In contrast, Large Language Models~(LLMs) are good at understanding complex natural language and generalizing across diverse contexts due to their pretraining on vast corpora.

% In the context of developer chatrooms, understanding question quality is crucial to improving their resolution outcomes. High-quality questions often include sufficient technical details, such as code snippets or error types, which make them easier to address. One way to extract such technical details is through Named Entity Recognition~(NER), a method used in information extraction that identifies and categorizes entities within the text, such as organizations and locations~\citep{li_ner_2022}. In software engineering, NER can be applied to identify software-specific entities, such as libraries, functions, and error messages. Previous research applied NER to Stack Overflow discussions to distinguish software-related entities from general text~\citep{ye2016software}. Leveraging NER in developer chatrooms allows for the extraction of technical details, providing a basis for evaluating question quality and understanding its relationship with resolution outcomes.

% Traditional NER techniques have proven effective in structured and formal text, they often require extensive feature engineering and task-specific training datasets, limiting their adaptability to informal, fragmented, and technical domains such as developer chatrooms. Large language models~(LLMs), by contrast, are good at understanding complex natural language and generalizing across diverse contexts due to their pretraining on vast corpora.

To address these challenges, we propose an approach named as \underline{\textbf{S}}oftwar\underline{\textbf{E}}-specific \underline{\textbf{N}}amed entity recognition, \underline{\textbf{I}}ntent detection, and \underline{\textbf{R}}esolution classification (SENIR) that leverages an LLM to label the chatroom conversations. SENIR recognizes software-specific entities, detects the intent behind a question in the context of developer chatrooms~\citep{huang_intent_2020}, and classifies whether a conversation has reached a satisfactory conclusion~(i.e., resolved) or remains unresolved. This study evaluates SENIR using the DISCO dataset~\cite{subash2022disco}, a collection of 29,243 developer chatroom conversations extracted from various channels related to software engineering~(SE) on Discord. Our work analyzes the developer chatroom conversations along the following three research questions (RQs):

\noindent\textbf{RQ1: \rqone}

Chatroom conversations are by nature informal, fast-paced, and lack sufficient context, making automated labelling a challenging task. To address this, we design SENIR using Mixtral 8x7B~\cite{jiang2024mixtral} to perform software-specific NER, intent detection, and resolution status classification. SENIR is evaluated on a manually labelled subset of 400 Discord conversations, achieving high accuracy (91\% for NER, 76\% for intent detection, and 93\% for resolution status), with corresponding F-scores of 86\%, 71\%, and 89\%.

\noindent\textbf{RQ2: \rqtwo}

We use SENIR to automatically label 29,243 developer conversations from the DISCO dataset to identify software-specific entities, intents, and resolution status. The labelled entities and intents from questions are used to engineer features together with additional features such as sentiment, posting time, and question length. These question-related features are used to train a mixed-effect model to predict resolution outcomes. The model achieves an AUC of 0.75 and the feature importance analysis reveals that positive sentiment and technical specificity~(e.g., use of library functions) positively impact resolution, while late posting times and excessive URLs negatively influence resolution. 

\noindent\textbf{RQ3: \rqthree}

We analyze the 29,243 labelled conversations to examine how entities and intents influence resolution success. Chi-Square tests reveal significant differences in resolution rates across intents, with \intent{API Usage} and \intent{API Change} demonstrate better resolution rates (33.6\% and 26.2\%) than \intent{Discrepancy} and \intent{Review} (22.0\%, and 18.8\%). Within intents, specific entity pairs show better resolution outcomes than others. For instance, the pair (\entity{Programming Language}, \entity{Library Function}) is particularly effective for technical intents like \intent{Errors}~(63.6\%), while (\entity{UI ELEMENT}, \entity{Website}) shows much lower success for intents like \intent{Discrepancy} and \intent{Review}. 

The main contributions of our paper are as follows:

\begin{enumerate}
\item We propose SENIR, an LLM-based approach to automatically label developer chatroom conversations with software-specific named entities, intents, and resolution status.

\item We provide a predictive model for predicting resolution status based on the questions posted and reveal the most influential features for a question to be resolved.

\item By investigating how specific entities and user intents influence question resolution, we provide actionable insights to help developers refine their questions to potentially achieve higher response rates.

\item We release the annotated developer conversation dataset which is derived from DISCO and augmented with the SENIR labels in our replication package~\cite{pouya2024replication}.
\end{enumerate}

The rest of the paper is organized as follows: Section~\ref{sec:approach} outlines our methodology, including the case study design and data analysis. Section~\ref{sec:results} presents the results for each research question. Section~\ref{sec:implications} discusses the implications of our findings. Section~\ref{sec:related_work} discusses related work. Section~\ref{sec:threats} discusses the threats to validity. Section~\ref{sec:conclusion} concludes the paper.

\section{Case Study Design} \label{sec:approach}

\begin{figure*}[!t]
\centering
    \includegraphics[width=.965\linewidth]{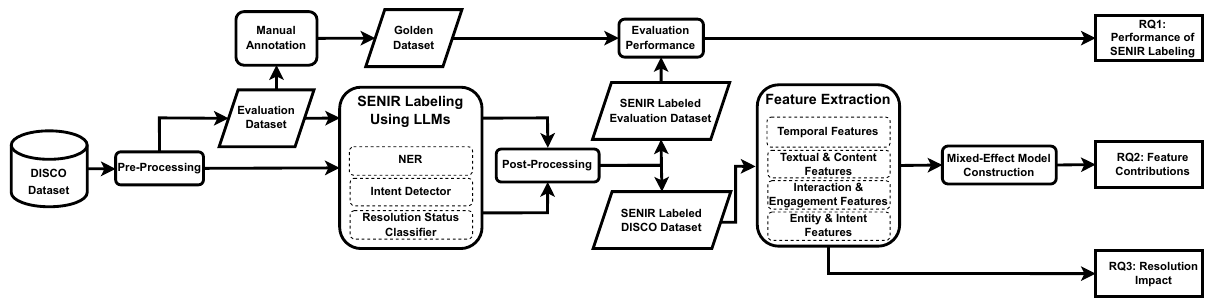}
\caption{Overview of our case study design.}
\label{fig:approach}
\vspace*{-1em}
\end{figure*}

This section presents our systematic approach for refining developer questions in chatrooms by leveraging an LLM. It includes details on data collection and preprocessing, our LLM-based approach for labelling developer chatroom conversations, manual labelling of a statistical sample set of conversations, and feature extraction. \figref{fig:approach} presents an overview of our approach.
% To systematically analyze chatroom conversations, we first collect and preprocess data from the DISCO dataset, ensuring that our study focuses on relevant developer discussions. We then apply SENIR, our LLM-based labelling approach, to extract software-specific entities, intents, and resolution statuses from these conversations. To validate the accuracy of our labelling process, we create a manually annotated golden dataset, which serves as a benchmark for evaluating SENIR’s performance. Finally, we extract a set of predictive features from the labelled dataset to investigate the factors influencing question resolution. These steps collectively enable us to assess SENIR’s effectiveness, develop a predictive model and gain insights into improving question formulation in developer chatrooms.

\subsection{Data Collection and Preprocessing}

To systematically analyze chatroom conversations, we first collect and preprocess data from the DISCO dataset~\cite{subash2022disco}, ensuring that our study focuses on relevant developer discussions. Table~\ref{tab:disco_channels} presents an overview of the dataset.

\begin{figure}[t]
\centering
    \includegraphics[width=.8\linewidth]{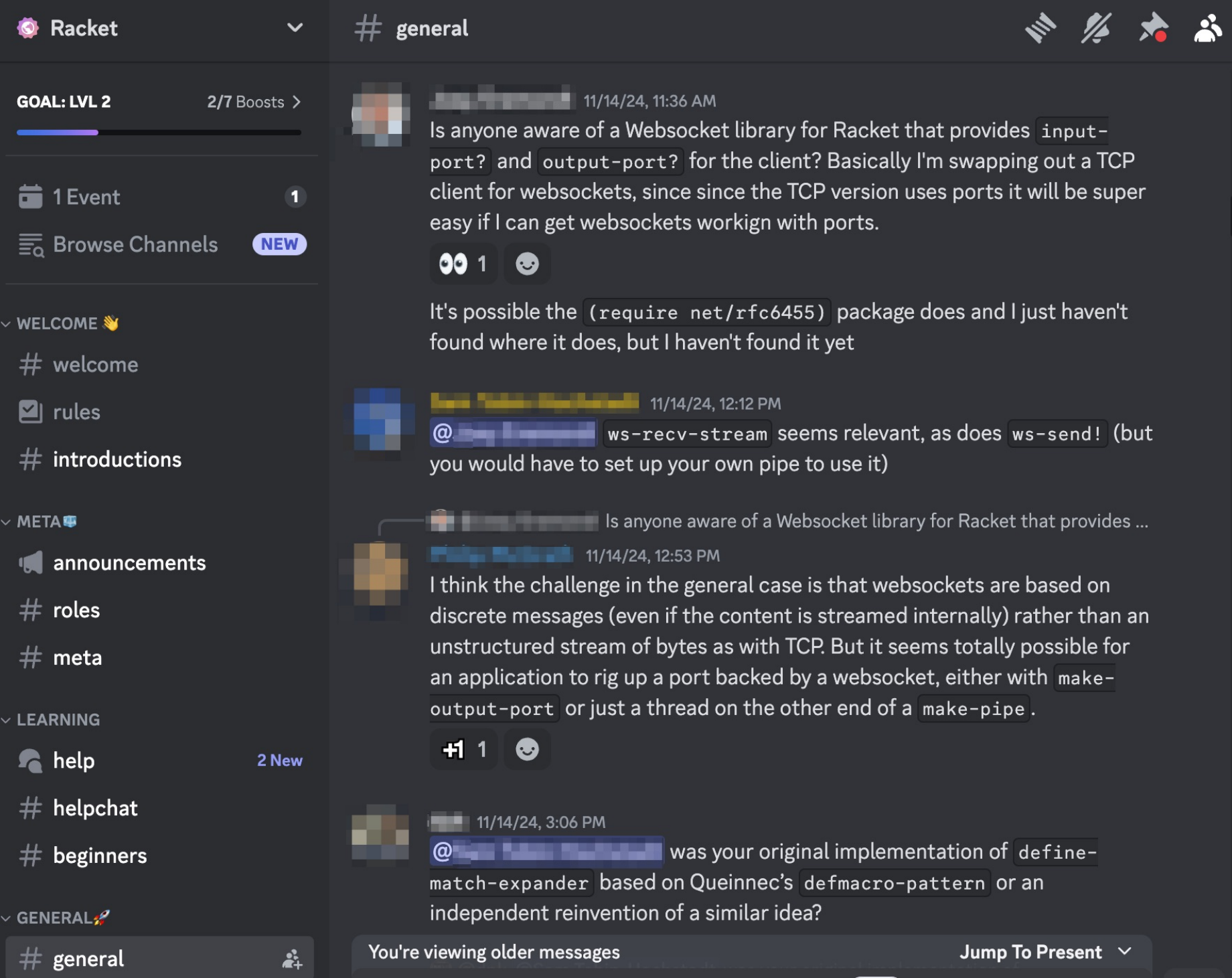}
\caption{A conversation in the \dquote{racket\#general} channel on Discord (The discussants' screen names are blurred for the purpose of privacy).}
\label{fig:conversation_example}
\vspace*{-1em}
\end{figure}

\begin{table}[t]
\centering
\caption{List of Channels in DISCO Dataset}
\label{tab:disco_channels}
\begin{tabular}{lr}
\toprule
Channel        & Number of Conversations \\
\midrule
python\#python-general      & 19,684                  \\
gophers\#golang              & 8,860                   \\
racket\#general & 538                    \\
clojurians\#clojure           & 161                    \\ \midrule
Total & 29,243 \\ \bottomrule
\end{tabular}
\vspace*{-1em}
\end{table}

\textbf{Step 1: Extracting messages and metadata.} We extract individual messages along with their associated conversation IDs, timestamps (\dquote{ts}), and user IDs from the DISCO dataset.  For instance, a message like \dquote{How do I install X?} might include metadata such as \dquote{Mia}~(the user who sent the message), \dquote{Aug 15, 2020, 10:34 AM}~(the time it was sent), and \dquote{12345}~(the conversation ID to which it belongs). \figref{fig:conversation_example} provides an example of a real conversation from the \dquote{racket\#general} channel of Discord.

\textbf{Step 2: Aggregating messages by conversation ID.} Since each message is recorded separately without direct references to other messages, we group messages sharing the same \dquote{conversation ID} to reconstruct complete interactions. This aggregation enables further analysis (e.g., resolution status classification) at the conversation level rather than the individual message level.

\textbf{Step 3: Augmenting conversations with additional metadata.} To better understand when and over what period a conversation took place, we augment each conversation record with additional metadata. For instance, we calculate the range of months over which a conversation took place (e.g., \dquote{Aug2020--Oct2020}) based on the start and end dates. This temporal information helps analyze patterns over time, such as how long discussions tend to last or identify specific periods of activity.

\begin{figure}[t]
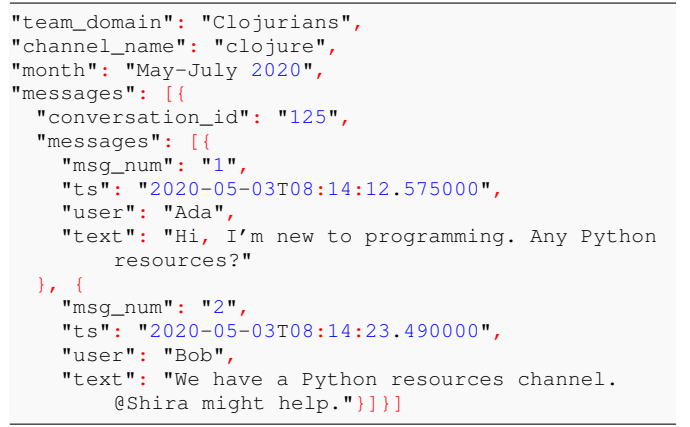

    \centering
\begin{lstlisting}[language=json, numbers=none]
"team_domain": "Clojurians",
"channel_name": "clojure",
"month": "May-July 2020",
"messages": [{
  "conversation_id": "125",
  "messages": [{
    "msg_num": "1",
    "ts": "2020-05-03T08:14:12.575000",
    "user": "Ada",
    "text": "Hi, I'm new to programming. Any Python resources?"
  }, {
    "msg_num": "2",
    "ts": "2020-05-03T08:14:23.490000",
    "user": "Bob",
    "text": "We have a Python resources channel. @Shira might help."}]}]
\end{lstlisting}
\caption{A sample JSON structure from the \dquote{clojurians\#clojure} channel.}
% \caption[A sample JSON code structure from the Clojurians channel.]
    % {A sample JSON code structure from the Clojurians channel.\footnotemark}
\label{fig:json_sample}
\vspace*{-1em}
\end{figure}

\figref{fig:json_sample} presents a sample JSON structure from the \dquote{clojurians\#clojure} channel. It contains various metadata fields, including \dquote{team domain}, \dquote{channel name}, \dquote{month}, \dquote{start date}, and \dquote{end date}, which provide context about where and when the conversations took place. Additionally, the \dquote{messages} field stores individual messages with attributes, such as \dquote{conversation ID}, \dquote{message number}, timestamp (\dquote{ts}), \dquote{user}, and \dquote{text}. %This structure helps to capture both the content of the conversation and the metadata required for detailed analysis, such as identifying user participation and the timeline of messages.

\subsection{SENIR for Labelling Developer Chatroom Conversations} \label{subsec:prompts}

We design SENIR to label developer chatroom conversations, focusing on three tasks: \textbf{(1)~Software-Specific NER:} NER identifies key software-related entities, such as \entity{Library}, \entity{Error Name}, and \entity{Version}, which are crucial for understanding the context of developer discussions. \textbf{(2)~Intent Detection:} Intent detection determines the purpose of each conversation, such as whether a user asks to learn a programming language or deals with an error. Understanding the intent of the initial question allows for better classification of conversations. \textbf{(3)~Resolution Status Classification:} This classification task assesses whether a conversation has successfully resolved the initial question. Classifying resolved and unresolved questions helps evaluate the effectiveness of the chatroom and identifies questions requiring further attention. Throughout this paper, we follow a consistent notation for entities and intents: 
entities are written in \texttt{typewriter} (e.g., \entity{Programming Language}), 
while intents are written in \textit{italics} (e.g., \intent{API Usage}).

% \begin{figure}[t]
%     \centering
% \begin{lstlisting}[language=json, numbers=none]
% [
%   {
%     "team_domain": "Clojurians",
%     "channel_name": "clojure",
%     "month": "May2020-July2020",
%     "start_date": "2020-05-06T05:35:13.419000",
%     "end_date": "2020-07-30T21:37:35.713000",
%     "messages": [
%       {
%         "conversation_id": "125",
%         "list of text": [
%           {
%             "message_number": "1",
%             "ts": "2020-05-03T08:14:12.575000",
%             "user": "Fidenciano",
%             "text": "Hi everyone, I'm new to discord and programming and looking for resources to learn python, Do you guys have any suggestions?"
%           },
%           {
%             "message_number": "2",
%             "ts": "2020-05-03T08:14:23.490000",
%             "user": "Gavriella",
%             "text": "Hi, we have a discord channel related to essential python resources. Maybe @Shira has some more advice for you."
%           }
%         ]
%       }
%     ]
%   }
% ]
% \end{lstlisting}

% \footnotetext{Clojurians Discord community: \url{https://discord.gg/5Vq8zDjt}}

\begin{table}
% \scriptsize
\centering
\caption{Software-Specific Entity Categories.}
\label{tab:entity_examples}
\begin{tabular}{p{2.4cm}p{3.8cm}p{1.3cm}}
\toprule
Entity               & Examples                                               & Reference \\ \midrule
\entity{Application}          & Mosh, JKplayer, api-java-client              & \cite{tabassum2020code,das2023zero,veera2019nerse,ye2016software}        \\ \midrule
\entity{Programming Language} & Python, Java, CSS, C++                                 & \cite{tabassum2020code,das2023zero,veera2019nerse,ye2016software}         \\ \midrule
\entity{Version}              & (Python) 2.7, (Windows) XP               & \cite{tabassum2020code,das2023zero}         \\ \midrule
\entity{Algorithm}            & UDP, DFS, RBM                                          & \cite{tabassum2020code}         \\ \midrule
\entity{Operation System}     & Linux, iOS, Windows                                    & \cite{tabassum2020code,das2023zero,veera2019nerse,ye2016software}         \\ \midrule
\entity{Device}               & Phone, Mobile, GPU                                     & \cite{tabassum2020code,das2023zero,veera2019nerse}         \\ \midrule
\entity{Error Name}           & Overflow, OutofRange, I/O Error                  & \cite{tabassum2020code}         \\ \midrule
\entity{User Name}            & John, Maya, Clark, @Maya                               & \cite{tabassum2020code,veera2019nerse}         \\ \midrule
\entity{Data Structure}       & Array, List, Hash table, Heap                   & \cite{tabassum2020code}         \\ \midrule
\entity{Data Type}            & String, Char, Double                                   & \cite{tabassum2020code}         \\ \midrule
\entity{Library}              & Numpy, Scipy, Auto-grad                                & \cite{tabassum2020code,veera2019nerse,ye2016software}         \\ \midrule
\entity{Library Class}        & ItemTemplate, actionManager     & \cite{tabassum2020code,veera2019nerse,ye2016software}         \\ \midrule
\entity{User Class}           & myClass, TestClass                                     & \cite{tabassum2020code,das2023zero}         \\ \midrule
\entity{Library Variable}     & math.inf, swing.color, ActionListener.Value            & \cite{tabassum2020code}         \\ \midrule
\entity{User Variable}        & user id, numberOfRowsInSection                         & \cite{tabassum2020code}         \\ \midrule
\entity{Library Function}     & numpy.isinf(), Math.floor()                            & \cite{tabassum2020code,veera2019nerse,ye2016software}         \\ \midrule
\entity{User Function Name}   & hello(), myFunction()                                  & \cite{tabassum2020code,veera2019nerse,ye2016software}         \\ \midrule
\entity{File Type}            & JSON, JAR                                              & \cite{tabassum2020code,veera2019nerse,ye2016software}         \\ \midrule
\entity{File Name}            & WindowsStoreProxy.xml, a.txt               & \cite{tabassum2020code}         \\ \midrule
\entity{UI Element}           & Button, Scroll bar, Text box                    & \cite{tabassum2020code}         \\ \midrule
\entity{Website}              & MSDN, Google, Yahoo                                    & \cite{tabassum2020code}         \\ \midrule
\entity{Organization}         & Apache, Microsoft Research, Fair                       & \cite{tabassum2020code,das2023zero}         \\ \midrule
\entity{License}              & CC BY 4.0, Apache 2.0
 & \cite{tabassum2020code,ye2016software}         \\ \midrule
\entity{HTML/XML Tag Name}    & h1, div, img                                           & \cite{tabassum2020code,veera2019nerse,ye2016software}         \\ \midrule
\entity{Value}                & \dquote{hello world}, 255, 30.5, True                 & \cite{tabassum2020code}         \\ \midrule
\entity{In Line Code}         & grep -rnw, select * from Tab                     & \cite{tabassum2020code}         \\ \midrule
\entity{Output Block}         & Output from console/any IDE                            & \cite{tabassum2020code}         \\ \midrule
\entity{Keyboard Input}       & CTRL+X, ALT, fn                                        & \cite{tabassum2020code}   \\ \bottomrule      
\end{tabular}
\vspace*{-1em}
\end{table}

\textbf{Step 1: Selecting the LLM.}
Although SENIR is compatible with various LLMs, we select Mixtral 8x7B~\cite{jiang2024mixtral} due to its top performance and larger context window~(33k tokens) compared to other open-source models on the chatbot arena leaderboard~\citep{chiang_llmarena_2024} as of January 23rd, 2024, when we started our research. The context window size refers to the maximum number of tokens that a model can process at one time when generating a response. A larger context window allows the model to take into account longer sequences of text, which is important when processing developer conversations that may contain up to 100 messages. 
% This capability enables SENIR to capture the full context of a conversation, including dependencies and nuances spread across multiple messages, thereby improving the quality of labelling.
% In the following steps, we describe the labelling process using the LLM:

\textbf{Step 2: Constructing prompts.}
To optimize performance, we design custom prompts to guide the LLM in handling specific tasks. The prompts include clear rules that instruct the model on what to focus on. For instance, in the NER prompt, the rules guide the LLM on which entities to extract~(e.g., \entity{Programming Language}). % To detect intents, we specify the rules that the model can use to identify the categories of intent~(e.g., \intent{Learning}, \intent{Review}, \intent{API Usage}). 
By following best practices in prompt engineering~\cite{openai_prompt, Chatcompletion}, we enhance the model’s ability to generate accurate and contextually relevant outputs.

\textbf{NER prompt example:} The model receives the initial question of a conversation along with timestamps, and is tasked with recognizing named entities. We collect a list of 28 software-specific entities based on prior work~\cite{tabassum2020code,das2023zero,veera2019nerse,ye2016software} and include this pre-defined list to guide entity recognition. Table~\ref{tab:entity_examples} provides examples and references for each of these entities. 

\vspace*{-0.5em}
\begin{center}
\begin{tcolorbox}[colframe=black!75!white, colback=gray!5, arc=2mm, boxsep=1.5mm, left=1.5mm, right=1.5mm, top=1mm, bottom=1mm, width=0.95\linewidth, breakable]
\footnotesize
\textbf{Prompt:} Extract all relevant software entities \textit{[list of entities]} from the text below.

\textbf{Input Format:}

Question: I am using TensorFlow version 2.5 and encountering an error during installation. How can I fix this?
% \begin{itemize}[leftmargin=*]
% \item User 1: I am using TensorFlow version 2.5 and encountering an error with installation.
% \item User 2: Have you tried updating the Python to version 3.8?
% \item User 3: It worked for me on Python 3.8 with TensorFlow 2.5.
% \end{itemize}

\textbf{Output Format:} 

[\dquote{TensorFlow: Library}, \dquote{2.5: Version}]
\end{tcolorbox}
\vspace*{-0.5em}
\end{center}
    
\textbf{Intent detection prompt example:} The model receives the initial question of a conversation and is tasked with detecting the intent. A list of 7 categories of pre-defined intents is used to label the conversation with one or more intents. The 7 categories are collected based on previous work~\cite{beyer2018automatically,rosen2016mobile,allamanis2013and,treude2011programmers,beyer2014manual,beyer2017analyzing} and Table~\ref{tab:intent_description} describes these intent categories. 

\vspace*{-0.5em}
\begin{center}
\begin{tcolorbox}[colframe=black!75!white, colback=gray!5, arc=2mm, boxsep=1.5mm, left=1.5mm, right=1.5mm, top=1mm, bottom=1mm, width=0.95\linewidth, breakable]
\footnotesize
\textbf{Prompt:} Identify the intents behind the question based on the following categories: [list of intents].

\textbf{Input Format:}

Question: How to install library X?
% \begin{itemize}[leftmargin=*]
% \item User 1: How to install library X?
% \item User 2: You need to use the command \`{}pip install library-x\`{}.
% \item User 1: Thanks, I am trying to learn how it works.
% \end{itemize}

\textbf{Output Format:} 

[\dquote{Learning}]
\end{tcolorbox}
\vspace*{-0.5em}
\end{center}

\textbf{Resolution status classification prompt example:} We also aim to use the LLM to classify whether a conversation is  \dquote{resolved} or \dquote{unresolved} based on the content of the full conversation.

\vspace*{-0.5em}
\begin{center}
\begin{tcolorbox}[colframe=black!75!white, colback=gray!5, arc=2mm, boxsep=1.5mm, left=1.5mm, right=1.5mm, top=1mm, bottom=1mm, width=0.95\linewidth, breakable]
\footnotesize
\textbf{Prompt:} Determine if the issue discussed in the following conversation is resolved.

\textbf{Input Format:}
\begin{itemize}[leftmargin=*]
\item User 1: How do I fix this bug in Library Y?
\item User 2: You should try reinstalling the dependency using version 3.0.
\item User 1: That fixed the issue, thanks!
\end{itemize}

\textbf{Output Format:} 

[\dquote{Resolved}]
\end{tcolorbox}
\vspace*{-0.5em}
\end{center}

\begin{table}
\centering
\caption{Descriptions of the Seven Intent Categories}
%\caption{Descriptions of the seven intent categories used in the conversation analysis\cite{beyer2018automatically}.}
\label{tab:intent_description}
% \begin{tblr}{
%   width = \linewidth,
%   colspec = {Q[140]Q[750]Q[100]}, % Adjusted the column widths to give "Intent" more space
%   hlines,
% }
\begin{tabular}{@{}p{1.2cm}p{5.5cm}p{1.15cm}@{}}
\toprule
Intent & Description                                                                                                                                                                                                                                                                                                                                              & Reference \\
\midrule
\intent{API Usage}       & This category subsumes questions of the types "How to implement something" and "Way of using something," as well as the categories "How-to" and "Interaction of API Classes." The posts in this category contain questions asking for suggestions on how to implement some functionality or how to use an API, with the questioner asking for concrete instructions. &   \cite{beyer2018automatically,rosen2016mobile,allamanis2013and,treude2011programmers,beyer2014manual,beyer2017analyzing} \\
\midrule
\intent{Discrepancy}     & This category contains questions about problems and unexpected behaviour of code snippets where the questioner has no clue how to solve them. It includes categories like "Do not work," "Discrepancy," "What is the Problem?" and "Why" (non-working code, errors, or unexpected behaviour).                                                                         & \cite{beyer2018automatically,rosen2016mobile,allamanis2013and,treude2011programmers,beyer2014manual}\\
\midrule
\intent{Errors}          & Posts in this category deal with the problems of exceptions and errors, often including an exception and the stack trace. It is equivalent to "Error and Exception Handling" and overlaps with "Why" (non-working code, errors, or unexpected behaviour).                                                                                                             & \cite{beyer2018automatically,rosen2016mobile,treude2011programmers,beyer2014manual,beyer2017analyzing}             \\
\midrule
\intent{Review}          & This category merges "Decision Help" and "Review," "Better Solution," and "What" (concepts), as well as "How/Why something works" (understanding, reading, explaining, and checking). Questioners ask for better solutions or reviews of their code snippets, best practices, or decision-making assistance.                                                        &        \cite{beyer2018automatically,rosen2016mobile,allamanis2013and,treude2011programmers,beyer2014manual}            \\
\midrule
\intent{Conceptual}      & This category includes questions about the limitations of an API, API behaviour, understanding concepts such as design patterns or architectural styles, and background information about some API functionality. It is equivalent to "Conceptual" and includes "Why?" and "Is it possible?" as well as "What" and "How/Why something works."                         &       \cite{beyer2018automatically,rosen2016mobile,allamanis2013and,treude2011programmers,beyer2014manual}             \\
\midrule
\intent{API Change}      & This category concerns questions arising from changes in an API or compatibility issues between different versions. It includes "Version" and "API Changes."                                                                                                                                                                                                         & \cite{beyer2018automatically,beyer2014manual,beyer2017analyzing}                  \\
\midrule
\intent{Learning}        & This category merges "Learning a Language/Technology" and "Tutorials/Documentation," where questioners seek documentation or tutorials to learn a tool or language on their own, rather than asking for specific solutions or instructions.                                                                                                                         &  \cite{beyer2018automatically,treude2011programmers,beyer2017analyzing}                \\ \bottomrule
\end{tabular}
\vspace*{-1em}
\end{table}

\begin{table*}
\centering
\footnotesize
\caption{The list of features used in the study and their descriptions.}
\label{tbl:feature_groups}
\begin{tabular}{llll}
\toprule
Category & Feature & Description & Reference \\ 
\midrule
\multirow{2}{*}{Temporal Features} 
    %& Duration & Total time span of the conversation & \makecell[tl]{\cite{ravi2014great,wise2011analyzing}} \\
    & Weekday & Day of the week the question took place & \makecell[tl]{\cite{ehsan2020empirical,anderson2012discovering}} \\
    & Daytime & Time of day the question occurred & \makecell[tl]{\cite{ehsan2020empirical}} \\
\midrule
\multirow{6}{*}{\begin{tabular}[c]{@{}l@{}}Textual \& \\ Content Features\end{tabular}}
    %& Number of Messages & Total number of messages in the conversation & \makecell[tl]{\cite{ravi2014great}} \\
    %& Average Message Length & Average length of messages & \makecell[tl]{\cite{calefato2018ask}} \\
    & Readability CLI & Readability score of the question text & \makecell[tl]{\cite{coleman1975computer,ehsan2020empirical,yang2016security}} \\
    %& Lexicons Conversation Length & Total number of words or lexicons & \makecell[tl]{\cite{ehsan2020empirical}} \\
    & Text-Code Ratio Question & Ratio of code snippets to regular text & \makecell[tl]{\cite{mondal2021early}} \\
    & URLs Count & Number of URLs mentioned & \makecell[tl]{\cite{ehsan2020empirical,vasilescu2014social}} \\
    & User Mentions & Number of times users are mentioned & \makecell[tl]{\cite{vasilescu2014social,backstrom2006group}} \\
    & Code Snippets & Presence of code blocks or snippets & \makecell[tl]{\cite{ehsan2020empirical,yang2016security}} \\
    & Question Length & Length of the initial question & \makecell[tl]{\cite{calefato2018ask}} \\
    %& Conversation Length & Total length of the conversation & \makecell[tl]{\cite{calefato2018ask}} \\
\midrule
\multirow{3}{*}{\begin{tabular}[c]{@{}l@{}}Interaction \& \\ Engagement Features\end{tabular}}
    %& Participants & Number of participants in the conversation & \makecell[tl]{\cite{ehsan2020empirical,mondal2021early,srba2016stack}} \\
    & Sentiment & Sentiment of the initial question & \makecell[tl]{\cite{hutto2014vader,guzman2013towards}} \\
    & Active Chatroom Questioner & Indicator of whether the questioner is an active member & \makecell[tl]{\cite{ehsan2020empirical,mondal2021early,srba2016stack}} \\
    & Questioner Received Response Ratio & Ratio of received responses to total questions asked & \makecell[tl]{\cite{mondal2021early,srba2016stack}} \\
\midrule
\multirow{6}{*}{\begin{tabular}[c]{@{}l@{}}Entity \& \\ Intent-Related \\ Features\end{tabular}}
    & Total Entities Count & Total number of entities recognized &  \\
    & Unique Entities Count & Number of unique entities &  \\
    & Entity Occurrences & Frequency of each entity within the question &  \\
    & Intent Total Count & Total count of identified intents &  \\
    & List of 7 Intents & Specific intent categories &  \\
    & List of 28 Entities & Specific software-related entities &  \\
\bottomrule
\end{tabular}
\vspace*{-1em}
\end{table*}

\subsection{Golden Dataset Collection and Manual Labelling}\label{subsec:golden_data}

We construct a golden dataset of software engineering-related chatroom conversations to evaluate the performance of the LLM in automatically labelling software-specific named entities, intent, and resolution status of a conversation. To ensure a representative sample, we randomly select a statistically significant subset of the collected conversations~(as listed in Table~\ref{tab:disco_channels}) with a confidence level greater than 95\% and a margin of error of 5\%. This results in a selection of 400 conversations randomly chosen from the dataset. We manually verify each conversation to ensure that it begins with a software engineering-related question.

The manual labelling process involves annotating conversations with software-related entities, intents, and resolution statuses. Two annotators, both are PhD students (one is the first author of the paper) in software engineering, independently annotate the dataset to ensure a broad and unbiased perspective. The annotation process is conducted in two phases to ensure consistency and minimize bias: 

\textbf{Phase 1: Initial annotation and agreement check.} The annotators first independently label a subset of 200 conversations, covering software-specific entities, intents, and resolution statuses. After that, we calculate inter-annotator agreement using Cohen’s Kappa~\citep{cohen2013statistical} to measure consistency in labels. The initial Kappa scores are 0.71 for entity recognition~(substantial agreement), 0.67 for intent detection~(substantial agreement), and 0.76 for resolution status~(substantial agreement). Disagreements, such as different interpretations of ambiguous intents like \intent{Conceptual}, are discussed, and labelling guidelines are refined to improve clarity and ensure alignment (e.g., specifying cues for distinguishing \intent{Learning} from \intent{API Usage}).

\textbf{Phase 2: Final annotation.} Using the refined guidelines from Phase 1, the annotators label the remaining 200 conversations. The final Cohen’s Kappa scores across all 400 conversations are 0.84 for entity recognition~(near-perfect agreement), 0.78 for intent detection~(substantial agreement), and 0.87 for resolution status~(near-perfect agreement).

The manual labelling process consists of the following steps:

\textbf{Step 1: Identifying software-related entities in the initial question of conversations.} 
%We identify and label software-related entities mentioned in the conversations. The labelling follows a NER approach, where we identify and tag occurrences of these entities. Each occurrence of the entity is labelled with the associated word and time stamp in the conversation.
Each annotator manually identifies and labels software-related entities within the initial questions of conversations. For each identified entity, the annotator records the exact word representing the entity (e.g., \entity{Library}) along with the timestamp of its occurrence in the conversation. This ensures precise tracking of where and when the entity appears in the discussion.

\textbf{Step 2: Categorizing intents in the initial question of conversations.} Each conversation is labelled based on its initial question with one or more intents that indicate its underlying purpose. The intent annotation follows pre-defined categories~(as shown in Table~\ref{tab:intent_description}) to maintain consistency.

\textbf{Step 3: Annotating resolution status for conversations.} 
Unlike Stack Overflow, Discord or similar chatrooms do not have a flag to indicate whether a question is resolved. To address this, each annotator manually labels the resolution status of each conversation. A conversation is labelled as \dquote{resolved} if the original question receives a relevant answer, and as \dquote{unresolved} if it does not.

% Disagreements are resolved through discussion, ensuring that the final dataset accurately reflects the consensus view of the annotators. To measure inter-annotator agreement, we use Cohen’s Kappa~\citep{cohen2013statistical}. Cohen’s Kappa measures the agreement between two annotators for categorical items. A higher Cohen’s Kappa score means a strong agreement between the annotators with a maximum possible value of 1 representing complete agreement~\cite{noei2024detecting}. A score above 0.80 is considered near-perfect agreement, while a score between 0.61 and 0.80 indicates substantial agreement. A score between 0.41 and 0.60 is interpreted as moderate agreement, and lower scores indicate fair to slight agreement. We achieve a Cohen’s Kappa score of 0.84 for entity recognition, indicating a near-perfect level of agreement, 0.78 for intent detection, indicating a substantial agreement and 0.87 for resolution status, indicating a near-perfect level of agreement.

% Keep it here in the experimental setup
\subsection{Feature Extraction}
\label{sec:mixed}
Feature extraction helps analyze and understand whether a conversation in the DISCO dataset can be resolved or not when the initial question is posted. The extracted features capture key aspects of the initial question, such as timing, content quality, interaction levels, and the presence of specific entities and intents. By linking these features to resolution outcomes, we gain insights into what makes a question more likely to be resolved. Feature extraction follows two steps:

\textbf{Step 1: Labelling the DISCO dataset using SENIR.} 
%To understand the dynamics within the conversations in the DISCO dataset, we leverage SENIR~(described in Section~\ref{subsec:prompts}) to label software-specific named entities, intents, and resolution statuses. 
We run SENIR on the entire dataset (illustrated in \figref{fig:approach}) which consists of 29,243 conversations to label the entities, intents, and resolution statuses for each conversation.

\textbf{Step 2: Extracting features from the labelled DISCO dataset.}
We engineer and extract a set of features from the labelled dataset and categorize them as temporal features, textual \& content features, interaction \& engagement features, and entity \& intent-related features. Each category and its related features are listed in Table~\ref{tbl:feature_groups} and described below. Except for entity \& intent-related features, all extracted features are collected from previous work~\citep{ehsan2020empirical,anderson2012discovering,mondal2021early,srba2016stack,coleman1975computer,yang2016security,vasilescu2014social,backstrom2006group,calefato2018ask,hutto2014vader,guzman2013towards}.

\textbf{Temporal features.}
These features capture the timing aspects of the initial question, based on the hypothesis that posting time may influence resolution likelihood. Specifically, we include \dquote{weekday}, representing the day of the week the question was asked, and \dquote{daytime}, indicating the hour of the day the question was posted (ranging from 0 to 23 hours). 

\textbf{Textual \& content features.}
This set of features is related to the structure and content quality of the initial question. The goal is to assess whether question clarity and readability impact resolution, as these can directly affect user engagement and the likelihood of achieving a resolution. For example, a higher readability score may indicate that the question is easier to understand. Features include \dquote{URLs count}, capturing the number of URLs present in the question, and \dquote{code snippets}, indicating whether the question contains code blocks.

\textbf{Interaction \& engagement features.}
These features focus on the questioner's activity, capturing how interaction and engagement may influence resolution likelihood. Examples include \dquote{active chatroom questioner}, which identifies whether the questioner is an active participant in the chatroom~(binary: 0 for inactive, 1 for active), and \dquote{questioner received response ratio}, which measures the proportion of the questioner's past questions that have received responses.

\textbf{Entity \& intent-related features.}
These features leverage the labelled outputs from SENIR to provide insights into the technical aspects of the initial question. For example, \dquote{total entities count} represents the total number of recognized entities in the question and \dquote{unique entities count} shows the number of distinct entities present. Additionally, \dquote{entity occurrences} measures how frequently entities appear within the question.

\section{Results}
\label{sec:results}

In this section, we provide the motivation, approach, and findings for each of our research questions.

\subsection{RQ1: \rqone} \label{sec:rq1}

\sectopic{Motivation.} 
Annotating chatroom conversations with relevant entities and intents can enhance problem-solving and knowledge sharing in SE communities like Discord. Precisely identifying key technical details can lead to higher-quality responses.
However, existing labelling approaches face significant challenges. Traditional NER and intent detection methods struggle with the informal, fragmented nature of developer chatrooms, where multiple topics often overlap. Manual labelling, while effective, is resource-intensive and limits scalability.

To address these challenges, we explore the use of the LLM to automate the labelling of developer chatroom conversations. With strong contextual awareness, the LLM chosen by our work may be able to better navigate informal discussions compared to traditional methods.
In this RQ, we evaluate the effectiveness of the LLM in automatically labelling software-related chatroom conversations. Specifically, we introduce SENIR, an approach that leverages the LLM to detect and label software-specific entities, user intents, and resolution status. The labelled dataset provides a basis for further investigation in RQ2 and RQ3.

\sectopic{Approach.} 
We evaluate the labelling performance of SENIR using the golden dataset~(see Section~\ref{subsec:golden_data}), which contains 400 manually annotated developer conversations. We compare labelling output from SENIR with manual labels using the performance metrics \emph{Precision}, \emph{Recall}, \emph{F-score}, and \emph{Accuracy}. \emph{Precision} measures the proportion of correctly predicted positive instances out of all predicted positives, while \emph{Recall} measures the proportion of true positives out of all actual positive instances. The \emph{F-score} represents the harmonic mean of \emph{Precision} and \emph{Recall}, balancing both metrics. \emph{Accuracy}, on the other hand, is the proportion of all correct predictions (both true positives and true negatives) to the total number of predictions~\cite{hand2012assessing}.

To calculate the evaluation metrics, we map SENIR’s predictions and the golden labels into the confusion matrix categories: \emph{True Positive} (TP), \emph{False Positive} (FP), \emph{False Negative} (FN), and \emph{True Negative} (TN). 

The evaluation metrics are calculated as follows:

\begin{equation} \text{Precision} = \frac{TP}{TP + FP} \end{equation}

\begin{equation} \text{Recall} = \frac{TP}{TP + FN} \end{equation}

\begin{equation} \text{F-Score} = \frac{2 \times \text{Precision} \times \text{Recall}}{\text{Precision} + \text{Recall}} \end{equation}

\begin{equation} 
\text{Accuracy} = \frac{TP + TN}{TP + FP + FN + TN} 
\end{equation}

\begin{table}[t]
\centering
\caption{Performance metrics for entity recognition, intent identification, and resolution status detection.}
\label{tab:performance_metrics}
\begin{tabular}{lrrr}
\toprule
Metric     & Entity (\%) & Intent (\%) & Resolution (\%) \\ \midrule
Accuracy   & 91          & 76          & 93              \\
Precision  & 88          & 72          & 96              \\
Recall     & 85          & 70          & 87              \\
F-Score    & 86          & 71          & 89              \\ \bottomrule
\end{tabular}
\vspace*{-1em}
\end{table}

\smallskip
The definition of each confusion matrix category depends on the specific labelling task:

\noindent{\textbf{(1) Entity Recognition (Token-Level)}}
\begin{itemize}[leftmargin=*]
    \item \textbf{Golden labels:} For each token in the initial question of a conversation, the golden dataset indicates either a specific entity type (e.g., \entity{Programming Language}, \entity{Error Name}) or non-entity.
    \item \textbf{Predicted labels:} SENIR assigns an entity type or non-entity to each token.
    \item \textbf{TP:} SENIR correctly labels a token with the same entity type as in the golden dataset.
    \item \textbf{FP:} SENIR assigns an entity type to a token that is labelled as non-entity or a different entity type in the golden dataset.
    \item \textbf{FN:} SENIR fails to assign an entity type to a token that should have one.
    \item \textbf{TN:} SENIR correctly labels a token as non-entity (or correctly refrains from assigning a wrong entity type).
\end{itemize}

\noindent{\textbf{{(2) Intent Detection (Conversation-Level, Multi-Label)}}
\begin{itemize}[leftmargin=*]
    \item \textbf{Golden labels:} Each conversation in the golden dataset may have one or more intent labels (e.g., \intent{API Usage}, \intent{Review}, \intent{Discrepancy}) based on the initial question.
    \item \textbf{Predicted labels:} SENIR outputs a set of intents for each conversation.
    \item \textbf{TP:} An intent predicted by SENIR matches one in the golden labels.
    \item \textbf{FP:} An intent predicted by SENIR does not appear in the golden labels.
    \item \textbf{FN:} An intent present in the golden labels is not detected by SENIR.
    \item \textbf{TN:} An intent is correctly not predicted (i.e., it appears in neither SENIR’s output nor the golden labels).
\end{itemize}

\noindent{\textbf{{(3) Resolution Status (Conversation-Level, Single-Label)}}
\begin{itemize}[leftmargin=*]
    \item \textbf{Golden label:} Each conversation is labelled with a single resolution outcome (i.e., \dquote{resolved} or \dquote{unresolved}).
    \item \textbf{Predicted label:} SENIR outputs exactly one resolution outcome for each conversation.
    \item \textbf{TP:} SENIR correctly predicts the outcome as \dquote{resolved} when the golden label is \dquote{resolved}.
    \item \textbf{FP:} SENIR predicts \dquote{resolved} when the golden label is \dquote{unresolved}.
    \item \textbf{FN:} SENIR predicts \dquote{unresolved} when the golden label is \dquote{resolved}.
    \item \textbf{TN:} SENIR correctly predicts \dquote{unresolved} when the golden label is \dquote{unresolved}.
\end{itemize}

\sectopic{Results.} 
\textbf{SENIR can correctly label 91\% of entities.} Table~\ref{tab:performance_metrics} summarizes of the performance metrics for each task, SENIR achieves strong performance in labelling resolution status with an F-score of 89\%. For the NER and intent detection tasks, SENIR also performs well with F-scores of 86\% and 71\%, respectively. Although the recall values are slightly lower, the precision and accuracy metrics remain robust, which indicates that SENIR can accurately capture the software-specific entities and intents in developer conversations.

\textbf{Our LLM-based approach shows strong performance for concrete entities (e.g., \entity{Error Name}, \entity{Library Function}), but lower performance for abstract ones (e.g., \entity{Application}).} Concrete entities, such as \entity{Error Name}~(94\% accuracy) and \entity{Library Function} entities~(88\% accuracy), are easier to detect likely due to their well-defined context. In contrast, abstract terms like \entity{Application} achieve a lower accuracy of 81\%. These results suggest that entities grounded in concrete, well-defined terms are easier for the model to identify compared to abstract concepts. This observation aligns with previous findings by Ye~\etal~\cite{ye2016software}, where the baseline approach also performs well on programming languages but struggles with more nuanced categories such as the \dquote{API category}.

\begin{figure}[t]
\centering
\includegraphics[width=\linewidth]{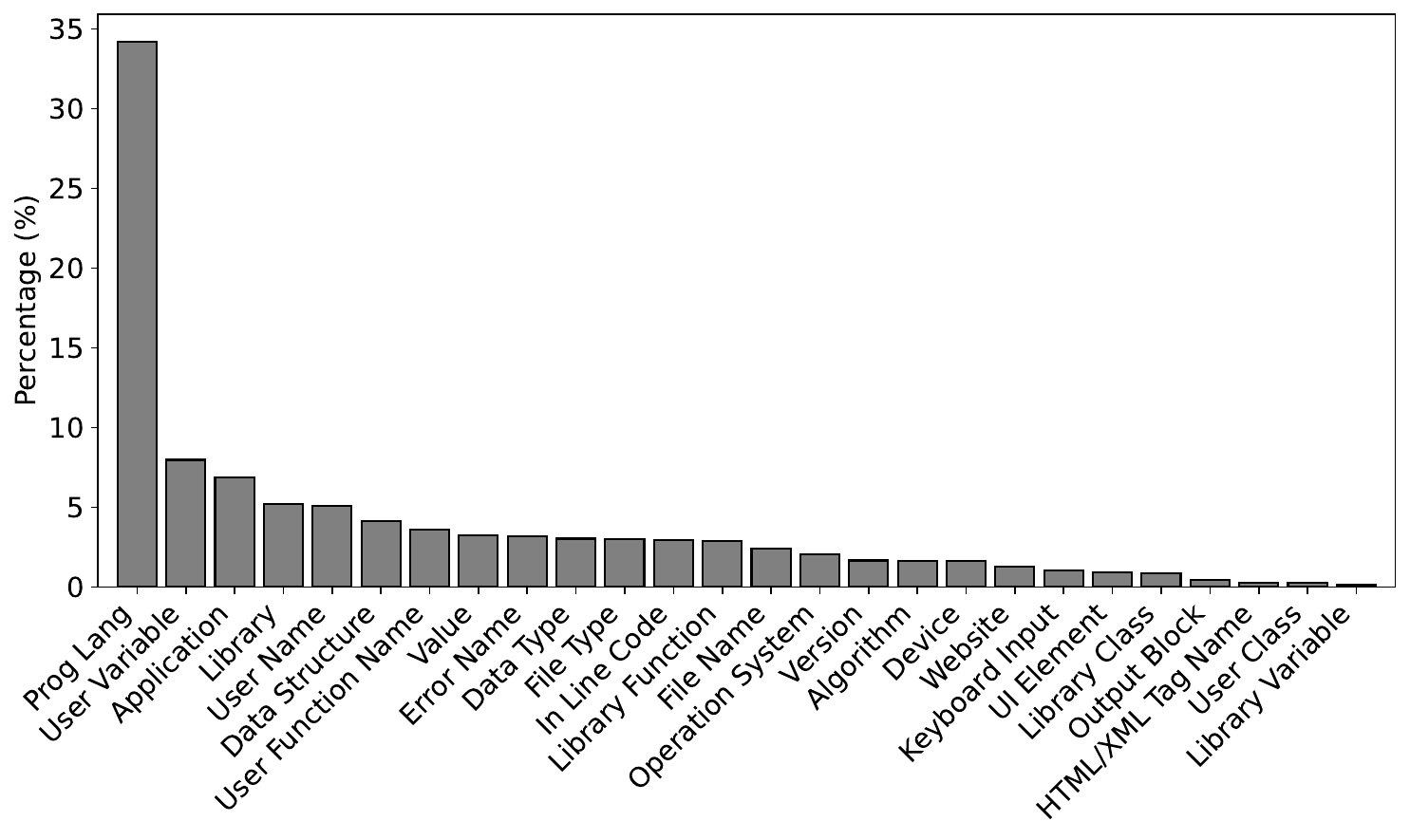}
\caption{Distribution of SENIR-labelled entities in the DISCO dataset.}
\label{fig:entities_bar_chart}
\vspace*{-1em}
\end{figure}

\textbf{Entities related to structured data (e.g., \entity{Version} and \entity{File Type}) consistently achieve high precision and recall.} Entities that represent structured information, such as \entity{Version} and \entity{File Type}, consistently rank the highest in terms of precision and recall~(as shown in \figref{fig:entities_bar_chart}). In contrast, entities such as \entity{Library} and \entity{In Line Code} show lower performance, which indicates that these categories may be more ambiguous or challenging for the model to detect in informal chatroom conversations.

\textbf{\entity{Programming Language} is the most frequently recognized entity in developer conversations.} \figref{fig:entities_bar_chart} shows the distribution of recognized entities, with \entity{Programming Language} appearing in 34\% of the conversations. This finding highlights the centrality of discussions surrounding programming languages in developer chatrooms. \entity{User Variable} and \entity{Application} are the next most frequent entities, indicating a significant focus on user-defined variables and software applications. 

\textbf{Intent detection shows varied performance, indicating differing levels of contextual complexity among intents.}
Table~\ref{tab:performance_metrics} shows that certain intents, such as \intent{Review} and \intent{Errors}, achieve high accuracy~(89\%), compared to more abstract intents, such as \intent{Conceptual}~(57\% accuracy). This discrepancy suggests that intents that require deep contextual understanding are more challenging for the model to capture accurately. In some instances, errors occur when entities or intents are used in unexpected contexts. For example, the term \squote{Go} is misclassified in discussions about travel rather than programming. This highlights a potential limitation in the model's ability to disambiguate between terms with multiple interpretations, particularly in casual or off-topic conversations.

\begin{figure}[t]
\centering\mbox{\hspace*{-2em}}
    \includegraphics[width=\linewidth]{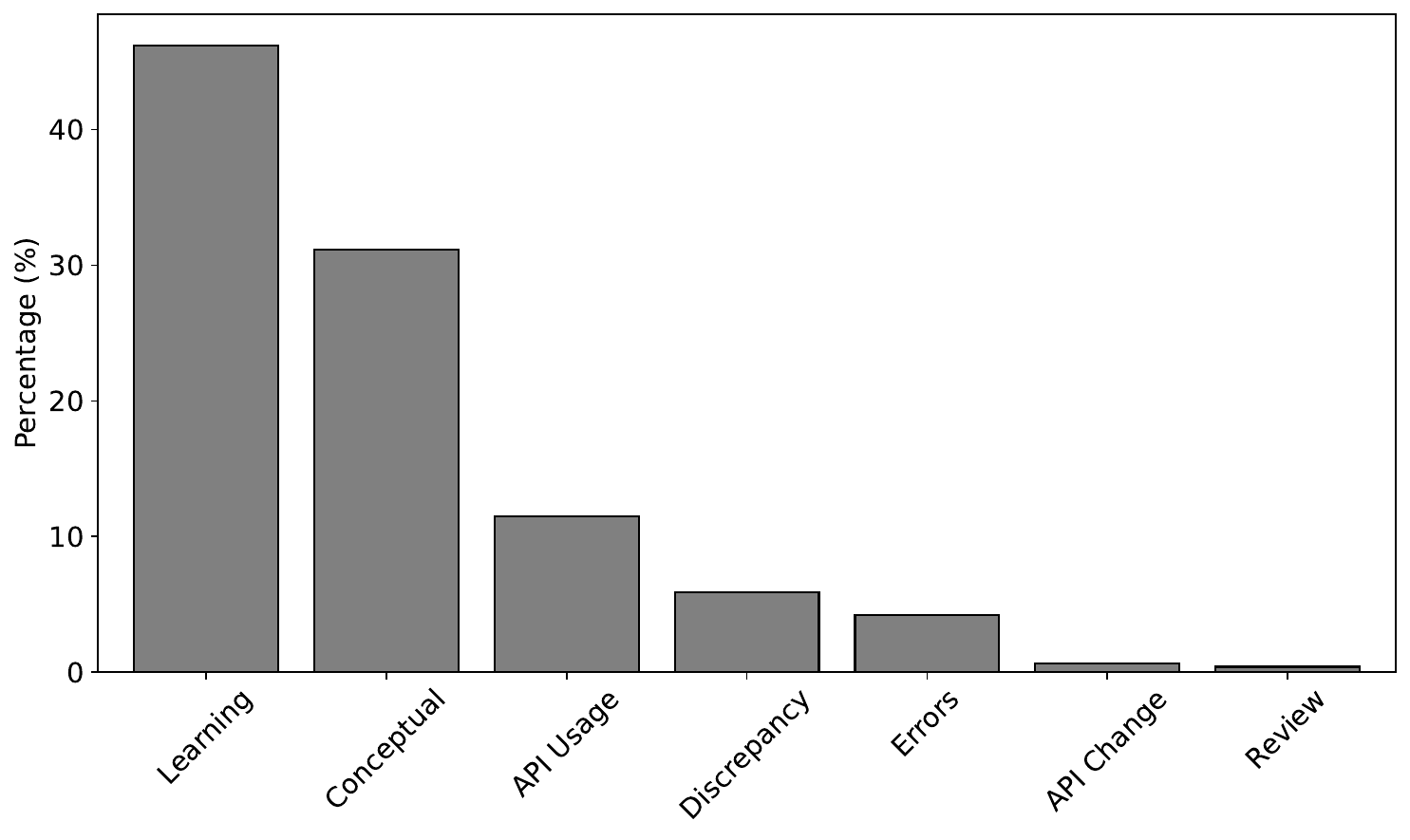}

\caption{Distribution of SENIR-labelled intents in the DISCO dataset.}
\label{fig:intents_pie_chart}
\vspace*{-1em}
\end{figure}

\textbf{Developer conversations predominantly focus on learning new concepts and understanding complex ideas.} \figref{fig:intents_pie_chart} shows that \intent{Learning} and \intent{Conceptual} are the most common intents in the dataset, representing 46\% and 31\% of all conversations. This reflects the focus of technical discussions on technical education and conceptual exploration. In contrast, \intent{API Usage}, the third most frequent intent~(11\%) highlights the need for clarifications or advice on API functionalities.

\textbf{Our LLM-based labelling approach shows strong agreement with manual labels.} Our approach achieves Cohen’s Kappa scores of 0.84 for entity recognition, 0.78 for intent detection, and 0.86 for resolution status. These scores fall within the \dquote{substantial} (0.61--0.80) and the \dquote{perfect} (0.81--1.00) agreement~\cite{mchugh2012interrater}, indicating that SENIR reliably captures key aspects of conversations. However, instances of disagreement between annotators and the model often arise due to conversational ambiguity. As the example shown below, certain conversations could be labelled as either \intent{Learning} or \intent{API Usage}, leading to different labels. Disagreement also occurs when users use vague terminology in a particular development community. These challenges highlight the difficulty of intent detection in casual, context-shifting environments like developer chatrooms. 
% An example of such ambiguity is illustrated below, where the discussion could be interpreted as both a learning-related inquiry and an API-related conversation.

\begin{center}
\begin{tcolorbox}[colframe=black!75!white, colback=gray!5, arc=2mm, boxsep=1.5mm, left=1.5mm, right=1.5mm, top=1mm, bottom=1mm, width=0.95\linewidth, breakable]
\footnotesize
% \textbf{Conversation:}
\begin{itemize}[leftmargin=*]
\item User 1: Hey folks, I’m confused about when to use async/await over threading in Python. Any insights?
\item User 2: Async is for concurrency, great for I/O-bound tasks. Threading is more about parallelism.
\item User 3: What’s your use case?
\item User 1: I have a function that fetches data from multiple APIs. Should I go with async or just use threads?
\item User 4: If it’s API calls, async is usually better. Use 
        `asyncio` and `aiohttp` for handling multiple requests.
\item User 1: Got it! I’ll try that. Thanks!
\end{itemize}
\end{tcolorbox}
\end{center}

\begin{Summary}{Summary of RQ1}{}
SENIR demonstrates high effectiveness in analyzing developer chatroom conversations by accurately identifying software-specific entities, intents, and resolution statuses, with an F-score of 86\% for entity recognition and 89\% for resolution status classification. Despite these strengths, challenges remain in detecting certain intents requiring deep contextual understanding, particularly for abstract concepts, highlighting areas for future improvement. 
%The results indicate that SENIR can significantly enhance the quality of automated analysis of developer chatroom conversations, providing actionable insights to improve question clarity and increase the likelihood of responses.
\end{Summary}

\subsection{RQ2: \rqtwo} \label{sec:rq2}

\sectopic{Motivation.}
In RQ1, we propose an LLM-based approach to label chatroom conversations with software-specific entities, the intent of the question, and the resolution outcome of the conversation. Building on this, we apply our approach to label a large dataset of 29,243 conversations sourced from Discord. This labelled dataset is used to further analyze the relationship between the characteristics of questions and their associated resolution outcomes.
The goal of RQ2 is to investigate which features of developer questions most significantly influence their likelihood of being resolved. 
Our goal is to gain insights into the factors that contribute to successful resolutions, as well as those that may harm them, by training a model on the question-derived features.

\sectopic{Approach.}  
To address RQ2, we follow the steps listed below:

\textbf{1. Labelling the Conversations}: We use the approach from RQ1 to label 29,243 conversations sourced from four distinct Discord channels within the DISCO dataset (see Table~\ref{tab:disco_channels}). Each conversation is annotated with software-specific entities, the intent of the question, and the resolution outcome. The resolution outcome is used as the dependent variable in building the prediction model. 

\textbf{2. Feature Extraction}: For each conversation, we isolate the initiating question and extract a comprehensive set of features from the initial questions, as described in Section~\ref{sec:mixed}. These features include general attributes, such as question length and sentiment, as well as features derived from the labelling in Step 1, such as entities count in a given question. Notably, the features pertain solely to the question itself. The model is not trained on any features irrelevant to the initial question as a whole. The feature set consists of 50 question-related features, which are listed in Table~\ref{tbl:feature_groups}.

\textbf{3. Feature Processing}: 
We first apply max-min normalization to scale all features to a range of [0, 1]. Without normalization, features with larger ranges (e.g., [-100, 100]) could disproportionately influence the model's performance compared to features with smaller ranges (e.g., [0, 10]). By transforming all features into the same range, we avoid potential biases during model training. 
The normalization formula is as follows:

\[
X_{\text{Normalized Value}} = \frac{X - X_{\text{min}}}{X_{\text{max}} - X_{\text{min}}} \in [0, 1]
\]

Where $X_{\text{min}}$ and $X_{\text{max}}$ are the minimum and maximum values of the feature, respectively.

After normalization, we perform a correlation and redundancy analysis to identify the highly correlated and redundant features.

\begin{figure}
    \centering
    \includegraphics[width=\linewidth]{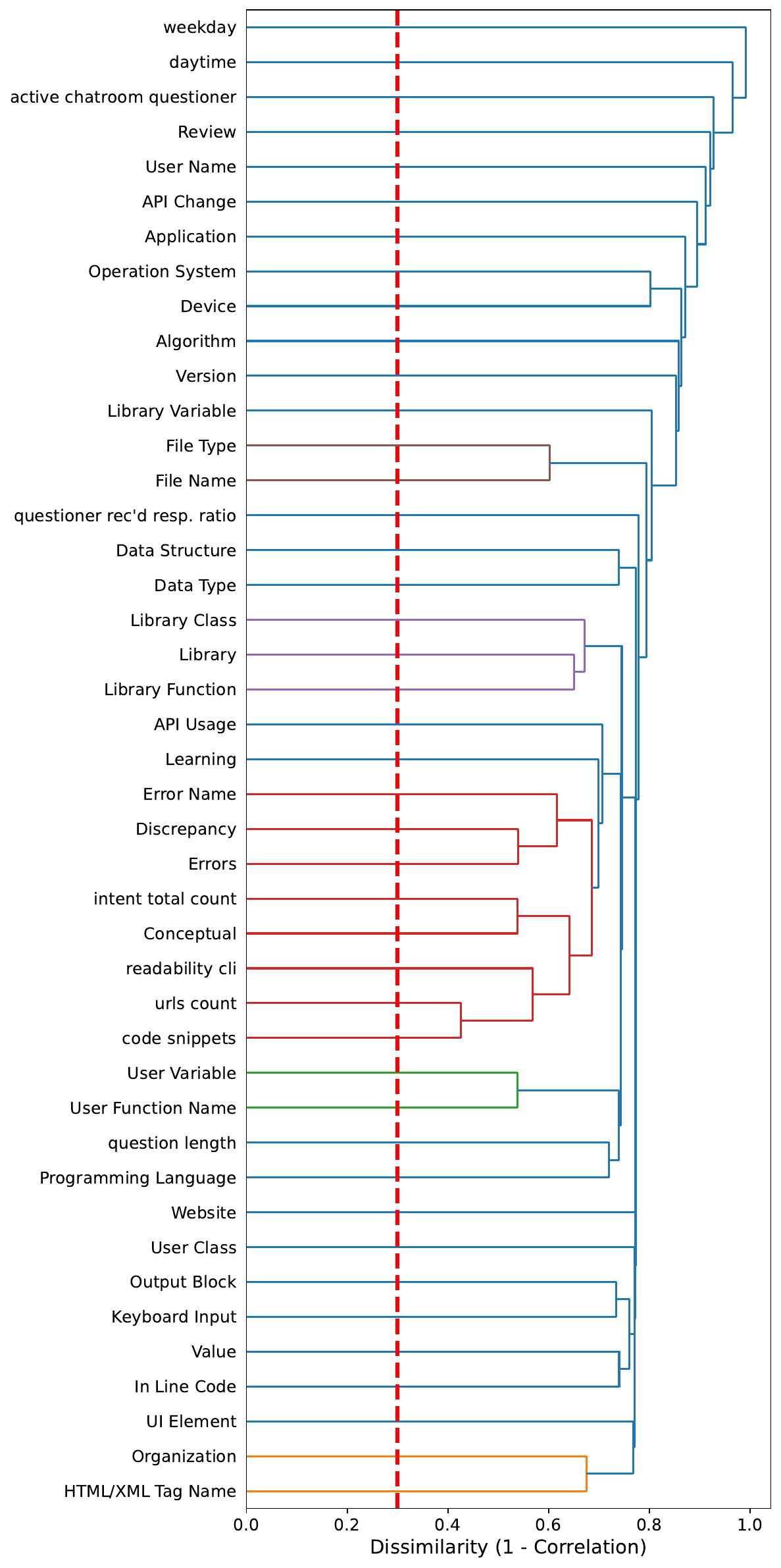} 
    \caption{Features correlation matrix dendrogram.}
    \label{fig:correlation_matrix_dendrogram}
\vspace*{-1em}
\end{figure}

\textbf{Correlation} refers to the degree to which two features are related. Highly correlated features provide similar information to the model, which can introduce instability in model coefficients and reduce interpretability. We use Spearman’s correlation coefficient to measure the degree of correlation between features. Following the previous studies~\cite{nguyen2010studying, noei2019too, noei2024detecting}, we consider a correlation coefficient threshold of 0.7 for the correlation coefficient values to identify strong correlations. To visualize the correlation structure, we apply hierarchical clustering, which requires a distance metric. We transform the correlation values into a dissimilarity measure defined as:

\begin{equation}
\text{distance} = 1 - \text{correlation}
\end{equation}

In this transformed space, highly correlated feature pairs (correlation 0.7) have distances 0.3. We therefore set a cut-off threshold of 0.3 on the (\figref{fig:correlation_matrix_dendrogram}) x-axis of the dendrogram and randomly remove one feature from each highly correlated pair to ensure that we retain a diverse and representative set of features.

\textbf{Redundancy} occurs when a feature does not provide unique information and can be predicted using other features. To detect redundancy, we compute the Variance Inflation Factor~(VIF), a widely used metric that quantifies how much a given feature is explained by other independent variables. Following prior work~\cite{pintas2021feature, jiarpakdee2016study}, we adopt a cut-off threshold of 10, where VIF values above this level indicate substantial redundancy. Features exceeding this threshold are excluded to improve computational efficiency and model generalizability.

After applying both correlation and redundancy filtering, we remove 15 highly correlated and redundant features: Text-Code Ratio Question, User Mentions, Total Entities Count, Unique Entities Count, Entity Occurrences, Sentiment, \entity{License}
Intent Total Count, \intent{API Usage}, \intent{Conceptual}, \intent{Discrepancy}, \intent{Errors}, \intent{Review}, \intent{API Chance}, and \intent{Learning}. This process reduces the feature space from 50 to 35 features while maintaining key predictive information.

\textbf{4. Model Training}: 
The dataset consists of a total of 29,182 questions, with 22,748 labelled as \dquote{unresolved}~(77.95\%) and 6,434 labelled as \dquote{resolved}~(22.05\%). This class imbalance is addressed through sampling strategies during model training.
To account for the variability of data sourced from different chatrooms, we train a mixed-effect model that incorporates:
\textbf{(i)} fixed effects, which capture the influence of extracted features, and 
\textbf{(ii)} random effects, which address variations across chatrooms.
To refine the model and prevent overfitting, we employ the stepwise regression algorithm~\citep{zhang2016variable}. Starting with an empty model, features are iteratively added based on their contribution to predictive performance. Specifically, we employ a stepwise feature selection approach, where features are evaluated based on their impact on performance metrics, such as the Akaike Information Criterion (AIC). Features that significantly improve model fit and predictive power are retained, while those with minimal or redundant contributions are excluded.  This process reduces the initial feature set to 20 impactful features, which are detailed in Table~\ref{tab:mixed_effects_model}.

Given the class imbalance, we explore two sampling strategies during training:
\textbf{(i)} Random undersampling, where the majority class is reduced to match the minority class, and
\textbf{(ii)} Oversampling using the Synthetic Minority Oversampling Technique (SMOTE), which generates synthetic samples for the minority class.

\textbf{5. LLM Baseline (Prompt-Based Classification)}: As a baseline for resolution prediction, we also prompt a general-purpose LLM, specifically Mixtral 8x7B~\cite{jiang2024mixtral}. 
In a zero-shot setting, we provide only the initial question to the LLM and ask it to predict whether the question is likely to be \dquote{resolved} or \dquote{unresolved}. 
An example of the prompt is shown below. 

\begin{center}
\begin{tcolorbox}[colframe=black!75!white, colback=gray!5, arc=2mm, boxsep=1.5mm, left=1.5mm, right=1.5mm, top=1mm, bottom=1mm, width=0.95\linewidth,breakable]
\footnotesize
\textbf{Prompt:} 
You are an AI assistant who helps analyze software engineering chatroom discussions. 
Your task is to determine whether the following developer question will likely receive a resolved answer based on its clarity, completeness, and technical details.

\textbf{Instructions:}
\begin{itemize}[leftmargin=*]
  \item Read the developer question carefully.
  \item Predict whether the question will be resolved (\textit{Yes}) or not resolved (\textit{No}).
  \item Provide a confidence score between 0\% and 100\%, indicating how sure you are about your prediction.
  \item Do not provide an explanation---only return the classification and confidence score.
\end{itemize}

\textbf{Input Format:}

Question: I am using TensorFlow version 2.5 and encountering an error during installation. How can I fix this?

\textbf{Output Format (Strict JSON format):}

\{\dquote{resolution\_status}: \dquote{Yes},
  \dquote{confidence\_score}: \dquote{82\%}\}
% \begin{verbatim}
% { "resolution_status": "Yes",
%   "confidence_score": "82%" }
% \end{verbatim}
\end{tcolorbox}
\end{center}

We then compare these baseline predictions (derived solely from the initial question) to the ground-truth resolution labels, which are obtained via an LLM-based method that has access to the \emph{entire} conversation (as explained in RQ1).

\textbf{6. Model Evaluation}: The trained mixed-effect model is evaluated using the \textbf{Area Under the Curve (AUC)}, which measures the model's ability to distinguish between resolved and unresolved questions. To ensure a stable evaluation, we use the following approaches:
\textbf{(i)} \textbf{5-Fold Cross-Validation~(CV)}, where the dataset is split into 5 folds, and each fold is used as a test set once; 
\textbf{(ii)} \textbf{10-Fold Cross-Validation~(CV)}, which follows the same approach except with 10 folds; and 
\textbf{(iii)} \textbf{Bootstrapping (100 iterations)}, where the dataset is repeatedly resampled with replacement to capture variability.
Each evaluation method is applied across both sampling strategies (random undersampling and SMOTE).

In parallel, we also evaluate the LLM’s baseline predictions to obtain an AUC value used for direct comparison.

\textbf{7. Feature Importance Analysis}: 
We conduct a feature importance analysis to identify which features have significant positive or negative impacts on the likelihood of resolution. We examine the coefficients from both the final model and the intermediary models at each step of the feature selection process. By analyzing the changes in coefficients as new features are added, we assess how each feature contributes to the model incrementally. Positive coefficients indicate features that increase the likelihood of resolution, while negative coefficients indicate those that decrease it. This stepwise examination ensures that the final set of features is robust and their impact remains consistent throughout the process.

To assess the statistical significance of these coefficients, we examine the associated \( z \)-values and \( p \)-values. The \( z \)-value measures how many standard deviations the coefficient is from zero under the null hypothesis ($H_0$: \textit{The feature does not affect the likelihood of question resolution}), with higher \( z \)-values indicating stronger evidence against the null hypothesis. Features with \( p < 0.05 \) are considered statistically significant.

\begin{table}[t]
\centering
\caption{AUC values of the mixed-effect models for the different configurations}
\label{tab:auc_results}
\begin{tabular}{lrrr}
\toprule
Configuration & 5-Fold CV & 10-Fold CV & Bootstrap \\ \midrule
Random Undersampling & 0.7544 & 0.7545 & 0.7531 \\
Oversampling with SMOTE & 0.7560 & 0.7558 & 0.7546 \\ 
\bottomrule
\end{tabular}
\vspace*{-1em}
\end{table}

\sectopic{Results.} 
\textbf{The model demonstrates acceptable discrimination performance across all configurations, with AUC values consistently falling within the range of 0.7 to 0.8.} Table~\ref{tab:auc_results} summarizes the AUC values for the tested configurations, which include two sampling strategies (random undersampling and oversampling with SMOTE) evaluated using three methodologies: 5-Fold CV, 10-Fold CV, and Bootstrapping.
Oversampling with SMOTE has a slight advantage and achieves the highest AUC values across all evaluation approaches, with an AUC of 0.7560 in 5-Fold CV and 0.7558 in 10-Fold CV. 
The stable results across all configurations provide confidence in the model's ability to distinguish between resolved and unresolved questions, allowing us to proceed with analyzing feature importance to understand the factors contributing to resolution outcomes.

Furthermore, when comparing these results to the LLM baseline, we observe that the AUC values for the LLM across different confidence scores remain consistently low, ranging from 0.50 to 0.53, indicating that its predictions are akin to random guessing.
We conclude that the performance of the baseline LLM falls well below the 0.70 to 0.80 range achieved by the feature-based mixed effect model. 
 
\begin{table}[t]
\centering
\caption{Mixed-Effect Model Results for Resolution Prediction}
\label{tab:mixed_effects_model}
\setlength{\tabcolsep}{2pt} 
\begin{tabular}{lrrrrcc}
\toprule
Feature & Coef. & Std.Err. & z & P>|z| & Sign. & Rel. \\
\midrule
%interaction\_diversity       & -0.613 & 0.015 & -42.085 & 0.000 & *** & $\searrow$ \\
%Posting Time                  & -0.196 & 0.015 & -12.821 & 0.000 & *** & $\searrow$ \\
Sentiment           & 0.705 & 0.044 & 15.912 & 0.000 & *** & $\nearrow$ \\
%API\_USAGE                   & 0.117 & 0.016 & 7.435  & 0.000 & *** & $\nearrow$ \\
\entity{User Name}                   & 0.097 & 0.017 & 5.745 & 0.000 & *** & $\nearrow$ \\
URLs Count                  & -0.298 & 0.077 & -3.846 & 0.000 & *** & $\searrow$ \\
%LEARNING                     & 0.080 & 0.016 & 4.881  & 0.000 & *** & $\nearrow$ \\
%ERRORS                       & 0.136 & 0.028 & 4.907  & 0.000 & *** & $\nearrow$ \\
\entity{Application}                  & -0.033 & 0.015 & -2.265 & 0.024 & * & $\searrow$ \\
%IN\_LINE\_CODE               & 0.050 & 0.010 & 4.885  & 0.000 & *** & $\nearrow$ \\
\entity{Library}                      & -0.062 & 0.018 & -3.472 & 0.001 & ** & $\searrow$ \\
%intent\_total\_count         & -20.746 & 5.657 & -3.667 & 0.000 & *** & $\searrow$ \\
Daytime                      & -0.020 & 0.014 & -1.475 & 0.140 &   & $\searrow$ \\
%ERROR\_NAME                  & -0.024 & 0.010 & -2.364 & 0.018 & * & $\searrow$ \\
\entity{Library Function}            & 0.065 & 0.022 & 2.931  & 0.003 & ** & $\nearrow$ \\
Weekday                      & 0.018 & 0.012 & 1.482  & 0.138 &   & $\nearrow$ \\
\entity{UI Element}                  & -0.061 & 0.043 & -1.415 & 0.157 &   & $\searrow$ \\
%question\_response\_time     & -0.443 & 0.217 & -2.046 & 0.041 & * & $\searrow$ \\
Code Snippets               & -0.380 & 0.103 & -3.705 & 0.000 & *** & $\searrow$ \\
\entity{Data Type}                   & 0.036 & 0.021 & 1.707  & 0.088 & . & $\nearrow$ \\
%OPERATION\_SYSTEM            & -0.021 & 0.011 & -1.971 & 0.049 & * & $\searrow$ \\
%DISCREPANCY                  & -0.043 & 0.023 & -1.838 & 0.066 & . & $\searrow$ \\
%REVIEW                       & 0.146 & 0.080 & 1.836  & 0.066 & . & $\nearrow$ \\
%API\_CHANGE                  & -0.085 & 0.053 & -1.596 & 0.110 &   & $\searrow$ \\
\entity{Data Structure}                   & 0.042 & 0.018 & 2.350  & 0.019 & * & $\nearrow$ \\
Readability CLI     & -0.347 & 0.216 & -1.605 & 0.109 &   & $\searrow$ \\
\entity{File Type}                   & -0.42 & 0.021 & -1.941  & 0.052 & . & $\searrow$ \\
\entity{Keyboard Input}                   & 0.84 & 0.037 & 2.241  & 0.025 & * & $\nearrow$ \\
\entity{Library Class}                   & 0.86 & 0.040 & 2.154  & 0.031 & * & $\nearrow$ \\
Question Length     & -0.124 & 0.072 & -1.716 & 0.086 & . & $\searrow$ \\
\entity{Organization}                   & -0.167 & 0.083 & -2.012  & 0.044 & * & $\searrow$ \\
Specific Intent Presence   & -0.115 & 0.016 & -6.976 & 0.000 & *** & $\searrow$ \\
\entity{HTML/XML Tag Name}                   & 0.124 & 0.069 & 1.782  & 0.075 & . & $\nearrow$ \\
%entity\_occurrences          & -0.120 & 0.029 & -4.193 & 0.000 & *** & $\searrow$ \\

\bottomrule
\end{tabular}
\vspace*{-1em}
\end{table}

\textbf{While certain SE entities provide helpful context to drive resolution, others may introduce unnecessary complexity or ambiguity.}
The subset of features representing SE entities—denoted in capital letters—provides insights into how the presence of specific entities influences question resolution, as shown in Table~\ref{tab:mixed_effects_model}. These features are binary, indicating whether a particular entity is present (1) or absent (0) in a question.
Among these, \entity{Application} (e.g., \dquote{Flask,} \dquote{PyCharm}), \entity{Library} (e.g., \dquote{NumPy,} \dquote{React}), and Code Snippets have a significant negative effect on resolution likelihood (\(p < 0.05\)). For instance, questions that include Code Snippets are negatively correlated with resolution (\(z=-3.705\), coefficient=-0.380), potentially indicating that such questions might introduce complexity or require additional context that impedes resolution. Similarly, the presence of \entity{Application} and \entity{Library} entities appears to slightly detract from resolution outcomes, possibly due to their broad or ambiguous nature.
In contrast, entities like \entity{Library Function} (e.g., \dquote{numpy.mean(),} \dquote{json.dumps()}) and \entity{Library Class} (e.g., \dquote{DataFrame,} \dquote{Button}) demonstrate a positive correlation with resolution (\(p < 0.05\)), suggesting that including specific, well-defined technical elements increases the likelihood of resolution. For example, the feature \entity{Library Function} has a coefficient of 0.065 (\(z = 2.931\)), indicating a modest but significant positive relationship. This finding highlights the value of technical specificity in driving resolution success.

Other features with a positive impact include Sentiment, where a more positive tone significantly improves resolution likelihood, and User Name, indicating that directly mentioning specific users could improve responsiveness. This result confirms the findings by~\citet{ehsan2020empirical}. On the other hand, URLs Count has a negative effect on resolution, indicating that later excessive URLs may introduce complexity or reduce focus, making questions harder to address effectively.

\begin{Summary}{Summary of RQ2}{}
 The mixed-effect model demonstrates acceptable performance in distinguishing resolved and unresolved questions, with stable AUC values across all configurations (0.7 to 0.8). Feature importance analysis highlights that while specific SE entities (e.g., \entity{Library Function}, \entity{Library Class}) and positive sentiment improve resolution likelihood, elements like excessive URLs, and entity mentions detract from resolution. 
\end{Summary}

\subsection{RQ3: \rqthree} \label{sec:rq3}

\sectopic{Motivation.}
The quality and formulation of developer questions are critical for improving response rates and overall quality of interactions. In RQ3, we aim to understand how specific combinations of software-specific entities and intents influence the question resolution. By providing insights into which entities and intents lead to higher resolution rates, our goal is to guide developers in crafting more effective questions, thereby reducing the number of unanswered questions.

\sectopic{Approach.}
We study the interaction among software-specific entities, intents, and the resolution of developer questions through the following aspects: 

\noindent\textbf{1. Intent Success Rate Evaluation:} To understand the overall efficacy of different question types in eliciting responses, we assess the success rate for each intent by calculating the percentage of resolved questions within each intent category. We investigate if there is a significant relationship between the intent of the question and its resolution status. To evaluate this, we use a Chi-Square test of independence~\cite{mchugh2013chi}, which determines if there is an association between two categorical variables. The test compares the observed frequencies~(the actual number of resolved and unresolved questions per intent) to the expected frequencies~(what would be expected if there were no relationship). A large Chi-Square statistic indicates that the observed and expected values differ significantly, suggesting a relationship between the variables. Conversely, a small Chi-Square statistic suggests little or no relationship. The significance of the test is evaluated by the $p$-value, where a small $p$-value (typically less than 0.05) indicates that the observed relationship is unlikely due to chance.

\noindent\textbf{2. Analysis of Entity Pairs:} We analyze entity pairs, defined as two entities appearing together in the initial question of a conversation. For questions involving three or more entities, all possible pairs are considered. For example, if a question contains the entities \entity{Device}, \entity{Application}, and \entity{Library}, the resulting pairs are (\entity{Device}, \entity{Application}), (\entity{Device}, \entity{Library}), and (\entity{Application}, \entity{Library}). This analysis is motivated by our observation that around 75\% of conversations involve at least two entities. The goal of this is to understand how these relationships contribute to question resolution. In addition, we examine the resolution success rates of entity pairs across different intent categories in their effectiveness. To ensure reliable results, we filter out pairs with fewer than 10 occurrences.

% \begin{table}
% \centering
% \caption{ Resolution Outcome by Intent }
% \label{tab:resolution_outcome_by_intent}

% \begin{tabular}{lllllll}
% Intent      & Success Rate & Number of Successful Combinations & Total Number of Combinations &  &  &   \\
% API Usage   & 46.91\%      & 629                               & 1341                         &  &  &   \\
% Review      & 47.22\%      & 17                                & 36                           &  &  &   \\
% Errors      & 37.15\%~     & ~ 175                             & 471                          &  &  &   \\
% Discrepancy & 32.47\%      & 200                               & 616                          &  &  &   \\
% Learning    & 37.94\%      & 1276                              & 3363                         &  &  &   \\
% Conceptual  & 33.80\%      & 974                               & 2882                         &  &  &   \\
% API Change  & 31.94\%      & 23                                & 72                           &  &  &   \\
%             &              &                                   &                              &  &  &   \\
%             &              &                                   &                              &  &  &   \\
%             &              &                                   &                              &  &  &  
% \end{tabular}
% \end{table}

\begin{table}[t]
\centering
\caption{Resolution Outcome by Intent}
\label{tab:resolution_outcome_by_intent}
\begin{tabular}{lrrr}
\toprule
Intent & \% Success & \# Success & \# Total \\ \midrule
\intent{API Usage} & 33.6 & 1,845 & 5,497 \\
\intent{API Change} & 26.2 & 81 & 309 \\
\intent{Errors} & 25.6 & 519 & 2,024 \\
\intent{Conceptual} & 23.7 & 3,535 & 14,924 \\
\intent{Learning} & 22.9 & 5,053 & 22,112 \\
\intent{Discrepancy} & 22.0 & 618 & 2,813 \\
\intent{Review} & 18.8 & 36 & 192 \\ \midrule
% No Intent & 15.8 & 172 & 1,086 \\ \midrule
Overall & 21.9 & 6,412 & 29,243 \\ \bottomrule
\end{tabular}
\vspace*{-1em}
\end{table}

\begin{table}[th]
\centering
\caption{Entity Pair Results (Top 3 and Bottom 3 by Success Rate).}
\label{tab:overall_intent_entity_combinations_success_rate}
\scriptsize
\begin{tabular}{@{}llrr@{}}
\toprule
Intent & Entity Pair & \% Succ & \# Total \\ \midrule
\multirow{6}{*}{\begin{tabular}[c]{@{}l@{}}Aggregate \\ Results\\ (All Intents)\end{tabular}} 
 & \entity{Device, Library Variable} & 45.5 & 11 \\
 & \entity{Version, Library Variable} & 41.7 & 12 \\
 & \entity{Data Type, Library Variable} & 36.4 & 22 \\ \cmidrule(l){2-4} 
 & \entity{Application, User Class} & 10.0 & 20 \\
 & \entity{User Variable, UI Element} & 8.4 & 154 \\
 & \entity{User Class, Value} & 6.9 & 29 \\ \midrule
\multicolumn{4}{l}{\textit{The following rows present entity pair results broken down by intent.}} \\ \midrule
\multirow{6}{*}{\intent{API Usage}} 
 & \entity{Application, File Type} & 53.3 & 15 \\
 & \entity{Operation System, User Variable} & 45.8 & 24 \\
 & \entity{Data Structure, Library Function} & 45.1 & 51 \\ \cmidrule(l){2-4} 
 & \entity{Value, Keyboard Input} & 5.6 & 18 \\
 & \entity{User Func Name, Keyboard Input} & 5.0 & 20 \\
 & \entity{Value, Output Block} & 0.0 & 10 \\ \midrule
\multirow{6}{*}{\intent{Conceptual}} 
 & \entity{Data Structure, Library Variable} & 42.1 & 19 \\
 & \entity{User Name, HTML/XML Tag Name} & 41.7 & 12 \\
 & \entity{Version, Library Variable} & 41.7 & 12 \\ \cmidrule(l){2-4} 
 & \entity{User Func Name, Website} & 6.9 & 29 \\
 & \entity{Algorithm, UI Element} & 6.3 & 16 \\
 & \entity{User Variable, Website} & 5.6 & 36 \\ \midrule
\multirow{6}{*}{\intent{Discrepancy}} 
 & \entity{User Variable, Library Function} & 39.1 & 115 \\
 & \entity{Library Class, Library Function} & 38.7 & 31 \\
 & \entity{Application, File Type} & 38.5 & 26 \\ \cmidrule(l){2-4} 
 & \entity{Data Type, Keyboard Input} & 0.0 & 10 \\
 & \entity{File Name, UI Element} & 0.0 & 11 \\
 & \entity{User Func Name, Keyboard Input} & 0.0 & 21 \\ \midrule
\multirow{6}{*}{\intent{Errors}} 
 & \entity{Prog Lang, Library Function} & 63.6 & 11 \\
 & \entity{Prog Lang, File Name} & 41.2 & 17 \\
 & \entity{Prog Lang, File Type} & 33.3 & 12 \\ \cmidrule(l){2-4} 
 & \entity{Prog Lang, Error Name} & 27.3 & 11 \\
 & \entity{Prog Lang, Version} & 14.3 & 14 \\
 & \entity{Prog Lang, User Name} & 0.0 & 10 \\ \midrule
\multirow{6}{*}{\intent{Learning}} 
 & \entity{Prog Lang, Library} & 53.9 & 13 \\
 & \entity{User Variable, User Func Name} & 50.0 & 10 \\
 & \entity{Prog Lang, User Func Name} & 46.2 & 13 \\ \cmidrule(l){2-4} 
 & \entity{Prog Lang, Application} & 15.4 & 13 \\
 & \entity{Prog Lang, Website} & 10.0 & 10 \\
 & \entity{Prog Lang, User Name} & 0.0 & 10 \\ \midrule
\multirow{6}{*}{\intent{Review}} 
 & \entity{Device, Library Variable} & 40.0 & 10 \\
 & \entity{Version, Data Structure} & 37.5 & 64 \\
 & \entity{UI Element, HTML/XML Tag Name} & 36.8 & 38 \\ \cmidrule(l){2-4} 
 & \entity{File Name, Website} & 10.0 & 90 \\
 & \entity{Application, User Func Name} & 9.1 & 154 \\
 & \entity{User Variable, UI Element} & 8.9 & 124 \\ \midrule
\multirow{6}{*}{\intent{API Change}} 
 & \entity{Data Structure, Output Block} & 75.0 & 12 \\
 & \entity{Algorithm, Error Name} & 66.7 & 12 \\
 & \entity{Data Type, UI Element} & 63.6 & 11 \\ \cmidrule(l){2-4} 
 & \entity{File Type, Keyboard Input} & 17.7 & 34 \\
 & \entity{Library, Website} & 17.1 & 35 \\
 & \entity{Application, User Func Name} & 17.0 & 59 \\ \bottomrule
\end{tabular}
\vspace*{-1em}
\end{table}

\sectopic{Results.}
\textbf{The success rates for the different intents vary considerably from 18.8\% to 33.6\%,  highlighting an overall suboptimal rate of resolution across all intents.} Table~\ref{tab:resolution_outcome_by_intent} shows that \intent{API Usage} (33.6\%) and \intent{API Change} (26.2\%) have the highest success rates, while \intent{Discrepancy} (22\%) and \intent{Review} (18.8\%) are on the lower end. This points to potential difficulties in addressing more complex or nuanced questions.
To further assess the relationship between intent and resolution status, a Chi-Square test is conducted. The test yields a Chi-Square statistic of 76.83 with 6 degrees of freedom and a $p$-value of $1.61e^{-14}$. The large Chi-Square statistic, far exceeding the critical threshold, and the small $p$-value strongly indicate that the differences in resolution rates across intents are statistically significant.
This suggests that certain intents are more likely to result in successful resolution than others, reinforcing the need for better question formulation in lower-performing categories like \intent{Discrepancy} and \intent{Review}.

\textbf{Entity pair analysis reveals that certain combinations are more effective in driving question resolution under specific intents.}
As shown in Table~\ref{tab:overall_intent_entity_combinations_success_rate}, for the \intent{API Change} intent, the pair (\entity{Data Structure}, \entity{Output Block}) achieves a success rate of 75\%, On the other hand, the pair (\entity{File Type}, \entity{Keyboard Input}) yields 17.7\% suggesting these entities might be less useful in eliciting responses.

For the \intent{API Usage} intent, the pair (\entity{Application}, \entity{File Type}) has a success rate of 53.3\%, showing that providing specific application-related and file-related information is beneficial. However, combinations involving the pair (\entity{Value}, \entity{Keyboard Input}) yield only 5.6\%, indicating their ineffectiveness in this context.

In the \intent{Learning} intent, the pair (\entity{Programming Language}, \entity{Library}) has a success rate of 53.9\%, while combinations involving the pair (\entity{Programming Language}, \entity{User Name}) results in a 0\% success rate, suggesting that providing detailed library information is much more beneficial than including user-specific information.

Regarding the \intent{Discrepancy} intent, we observe a shift towards diagnostic-focused entities such as \entity{Library Function} and \entity{User Variable}}. However, combinations involving the pair (\entity{UI Element}, \entity{File Name}) show lower success rates, indicating that backend-related information tends to be more effective for resolution.

% We observe that "Programming Language" and "User Variable" appear as dominant entities, frequently appearing in most developer conversations, regardless of the context. This entity combination consistently appears in intents such as "API Usage," "API Change," "Learning," and "Conceptual," indicating their importance in framing both practical and theoretical questions. 
%For more specific intents like "Errors" and "Discrepancy", we observe a shift towards diagnostic-focused entities such as "Error Name", "Library", and "In-Line Code".

Similarly, the \intent{Review} intent tends to occur with entities related to detailed codebase elements, such as \entity{Library Variable}, \entity{Version} and \entity{UI Element}, indicating that these conversations often involve reviewing or resolving problems within specific parts of the codebase.

Our observations show that the SENIR-labelled entities accurately capture the technical content of the questions and reflect the associated intents, as initially intended.

\begin{Summary}{Summary of RQ3}{}
The analysis in RQ3 reveals that specific combinations of software-specific entities and intents impact the likelihood of question resolution. Dominant entities such as \entity{Programming Language} and \entity{Library} play a key role across multiple intents, while entities like \entity{User Variable} and \entity{UI Element} are less beneficial. Success rates vary considerably among intents, with \intent{API Usage} and \intent{API Change} showing the highest resolution rates, while \intent{Discrepancy} and \intent{Review} show lower rates. The Chi-Square analysis confirms that resolution rates significantly differ across intents, suggesting that better question formulation is needed for lower-performing categories.

\end{Summary}

\section{Implications} \label{sec:implications}

In this section, we discuss our findings and their possible implications for developers, chatroom platforms, and researchers.

\subsection{Implications for Developers}

%\textbf{Developers should craft focused questions with specific technical details to improve resolution rates.} Our findings in Sections~\ref{sec:rq2} and~\ref{sec:rq3} highlight that precise entities, such as \entity{Library Function} or \entity{Library Class}, are strongly associated with higher resolution rates, particularly for intents like \intent{Discrepancy} and \intent{Errors}. Developers should prioritize clearly specifying technical details, such as the exact function or class being used, rather than relying on broad or vague descriptions. For instance, instead of including large \entity{CODE SNIPPETS} or abstract entities like \entity{Application}, developers should narrow their questions by providing concise and targeted information about the issue at hand.

\textbf{It is important to provide focused questions with specific technical details.} Our findings (Sections~\ref{sec:rq2} and~\ref{sec:rq3}) reveal that including precise entities, such as \entity{Library Function} or \entity{Library Class}, correlates with higher resolution rates, particularly for intents like \intent{Discrepancy} and \intent{Errors}. Rather than relying on large \entity{CODE SNIPPETS} or abstract references (e.g., \entity{Application}), developers should pinpoint the exact function or class in question.

%\textbf{Developers should adopt a positive tone and directly mention specific users to improve resolution rates.} Our findings in Sections~\ref{sec:rq2} reveal that questions written in a positive tone correlate with higher resolution rates, likely because they foster a more engaging and collaborative atmosphere. Additionally, directly mentioning specific users (e.g., \dquote{@UserName}) increases the likelihood of resolution by directly addressing experts or experienced contributors in the community. Developers can benefit from adopting a polite, appreciative tone and addressing their questions directly to relevant individuals when appropriate.

\textbf{It is valuable to maintain a positive tone, tag specific users, and avoid overloading with URLs.} 
As highlighted in Section~\ref{sec:rq2}, questions formulated with a positive tone are more likely to be resolved. Likewise, tagging specific users (e.g., \dquote{@UserName}), when appropriate, increases visibility and may improve the response rate. However, including too many URLs can overload the conversation and detract from the core question, which can potentially slow down or prevent the resolution.

\subsection{Implications for Chatroom Platforms}

%\textbf{Chatroom platforms should provide structured question templates to guide users in writing better questions.} Our findings show that questions that contain specific entities and intents have higher resolution rates~(Section~\ref{sec:rq3}). Chatroom platforms can leverage this insight by offering structured templates to guide users. For instance, if a user indicates that they are asking about an error, the platform could suggest a template prompting them to provide details about the \entity{Programming Language}, \entity{Library}, and relevant \entity{Library Function}. Such templates can reduce ambiguity and improve the quality of questions submitted to the platform.

\textbf{Chatroom platforms can benefit from offering structured question templates.} Questions that specify concrete entities and intents correlate with higher resolution rates (Section~\ref{sec:rq3}). Chatroom platforms can leverage these insights by providing structured templates that prompt users to include important technical details. For example, if a user indicates they are dealing with an error, the platform could prompt them to specify the \entity{Programming Language}, \entity{Library}, and relevant \entity{Library Function}, to help minimize ambiguity of the questions.

%\textbf{Chatroom platforms should integrate SENIR to automatically generate tags and highlight key entities to assist responders in reviewing questions.} The results in Section~\ref{sec:rq1} demonstrate that SENIR can successfully label software-specific entities and intents. Chatroom platforms can integrate SENIR to automatically generate tags for questions, highlighting key entities (e.g., \entity{Programming Language}) and intents (e.g., \intent{API Usage}). This would allow responders to quickly assess the context and filter out questions based on their expertise, improving response efficiency and accuracy.

\textbf{Chatroom platforms can benefit from integrating an approach like SENIR for automated tagging and highlighting}. As shown in Section~\ref{sec:rq1}, SENIR effectively labels software-specific entities and intents. Chatroom platforms could integrate SENIR to automatically generate tags for new questions to draw attention to key entities (e.g., \entity{Programming Language}) and identify intents (e.g., \intent{API Usage}). This approach can enable developers to quickly assess a question’s context and filter questions based on their expertise.

\subsection{Implications for Researchers}

\textbf{Researchers can use SENIR to label other developer conversation datasets.} The results in Section~\ref{sec:rq1} confirm that SENIR can reliably label developer conversations with software-specific entities, intents, and resolution statuses. While our study focuses on Discord, SENIR is generalizable and can be applied to other platforms, such as GitHub Discussions.\footnote{\url{https://docs.github.com/en/discussions}} Researchers can use SENIR to investigate different platforms and analyze how entities and intents influence resolution rates in various contexts. %opening up new opportunities for studying developer interactions and improving these platforms.

\textbf{Researchers can leverage SENIR to enhance chatbots in developer chatrooms.} SENIR’s entity, intent, and resolution labels provide valuable signals for improving chatbot-based assistance in chatrooms. For example, researchers can use the resolution status to curate high-quality resolved questions for retrieval-augmented generation~(RAG)~\cite{lewis2020rag}. This enables chatbots to provide more reliable answers based on past conversations. In addition, entities and intents can refine retrieval strategies, techniques such as GraphRAG~\cite{han2025graphrag} can build knowledge graphs to enhance retrieval by leveraging labelled entities and intents. 
%By incorporating SENIR’s labels, researchers can improve the accuracy and contextual relevance of chatbot responses, making chatbots more effective in assisting developers.

\section{Related Work}\label{sec:related_work}

In this section, we discuss related work about NER in software engineering contexts, LLM applications in software engineering tasks, and developer chatrooms and online forums.

\subsection{NER in Software Engineering}

NER has been widely studied in the software engineering domain for automatically identifying and categorizing software-specific entities within text from various sources such as source code, commit messages, documentation, and social media content. \citet{ye2016software} developed machine learning based NER systems for software engineering social content. Their approach demonstrates improved performance over rule based systems by addressing entity ambiguity and informal language. Similarly, \citet{tabassum2020code} highlighted the importance of domain-specific NER for understanding code-related discussions. Their approach emphasizes the role of software-specific categories like APIs, frameworks, and libraries.

Recent advancements leverage transformer models like BERT for NER tasks, which have improved contextual understanding in software engineering texts. SoftNER~\cite{tabassum2020code}, a BERT-based model, achieved notable success in identifying code tokens and software-related entities within Stack Overflow conversations. \citet{das2023zero} further explored zero-shot NER techniques to recognize unseen entities, showcasing the adaptability of pre-trained models in low-resource scenarios.
While these studies primarily focus on structured platforms like Stack Overflow, our work targets the fragmented and dynamic nature of developer chatrooms, such as Discord. By leveraging LLMs and integrating intent detection alongside NER, SENIR labels entities and intents to improve question clarity and resolution outcomes in unstructured environments.

\subsection{LLM Applications in Software Engineering}

LLMs have shown significant potential in automating a broad range of SE tasks, such as code generation, bug detection, and documentation~\cite{hou2024large, fan2023large}. Although several studies focus on retrieving Q\&A content from structured platforms~\cite{zhang2020chatbot4qr}, researchers have also explored LLM-based methods for tasks like requirements engineering~\cite{preda2024supporting,hassani2025empirical} and code refinement~\cite{guo2024exploring}. \citet{colavito2024leveraging} further demonstrated that GPT-like models can effectively classify GitHub issues, reducing human workload in issue labelling.

Our work extends LLM applications to dynamic chatroom environments by introducing automated labelling of entities and intents and leverages the deep contextual understanding of LLMs to enhance question clarity, enabling structured analysis and actionable feedback for developers. This integration extends the applicability of LLMs to dynamic chatroom environments, where conventional retrieval methods struggle.

\subsection{Developer Chatrooms and Online Forums}

Developer chatrooms (e.g., Slack, Discord) enable real-time collaboration, but their informal style complicates message analysis. Empirical studies have shown that missing details often lead to unanswered questions, whereas user mentions can boost engagement~\cite{ehsan2020empirical}. \citet{subash2022disco} introduced the DISCO dataset to highlight the unique challenges of disentangling and analyzing these multi-threaded discussions. Similarly, \citet{el2021exploring} and \citet{shi2021first} found that developer chatrooms contain valuable knowledge but remain difficult to study due to their fluid format.

\citet{lill2024helpfulness} found that reusing past chats and Q\&A posts to resolve new Discord questions is helpful in only 40\% of cases—partly because questions often lack clarity. Similarly, \citet{tufano2024unveiling} examined how developers use LLMs like ChatGPT to seek assistance in open-source projects, underscoring the growing influence of AI-based aids. While these studies focus on overall chatroom behaviour, our work aims to refine questions by jointly modelling NER, intent, and resolution status within a unified framework. By extracting software-specific entities and understanding the question’s underlying purpose, we enable structured insights that pave the way for higher-quality discussions and faster problem resolutions.

\section{Threats to Validity}\label{sec:threats}

In this section, we discuss the threats to validity of our study about refining developer questions in chatrooms.

\textbf{Construct Validity.}
We study only one LLM, Mixtral, due to its larger token size compared to other open-source models. However, this may limit the generalizability of our results to other models. To mitigate this threat, we carefully select a set of software-specific entities and intent categories that are widely recognized within the software development community. Two PhD students with expertise in software engineering and natural language processing manually label the dataset, ensuring that the entity and intent classification schemes accurately reflect real-world scenarios.

\textbf{Internal Validity.}
The primary threat is potential bias in manual labelling and the influence of specific prompt designs on the results.
To address this, we employ a rigorous labelling process with two annotators and calculate the inter-annotator reliability using Cohen’s Kappa to reduce subjective bias. Moreover, we test different prompt designs in a preliminary study to select the most effective ones for our main experiments.

\textbf{External Validity.}
Since our study focuses on Discord chatrooms, the results may not be directly applicable to other software engineering platforms such as Stack Overflow or GitHub Discussions. Furthermore, different communities may have varying norms and communication styles that could affect the performance of our approach. To mitigate this threat, we analyze a large dataset spanning multiple software engineering-related chatrooms to ensure our findings are not specific to a single community. We also select chatrooms with diverse programming topics to improve the applicability of our results. While we acknowledge the diversity in software development communities, we recommend further validation in other environments to ensure broader generalizability.

\section{Conclusion}\label{sec:conclusion}

In this study, we present SENIR, an LLM-based approach for labelling chatroom conversations with software-specific named entities, intents, and resolution outcomes to understand and refine developer questions. Through the lens of three research questions, we demonstrate how these structured labels deepen our understanding of developer Q\&A in chatrooms and guide improvements in question formulation. Our experiments on 29,243 conversations from the DISCO dataset showed SENIR’s robust performance for entity extraction (\textbf{86\% F-score}), intent detection (\textbf{71\% F-score}), and resolution status classification (\textbf{89\% F-score}). Leveraging these labels, we built predictive models of conversation resolution and found that certain entity-intent combinations (e.g., {\entity{Library Function}} with \intent{Errors}) increase success, while features like excessive URLs and late posting times hinder resolution. A Chi-Square analysis further confirmed significant differences in resolution rates across various intents, suggesting actionable paths for refining developer questions. Future research can build on our study by exploring real-time feedback mechanisms or extending the approach to additional developer support communities, thereby shaping more targeted, efficient, and high-resolution Q\&A.

% Acknowledgments
\section*{Acknowledgments}
We want to express our appreciation to Bihui Jin at the University of Waterloo for helping to manually validate our results in order to build the golden dataset. His commitment was really helpful in confirming the reliability and precision of our findings.

% Bibliography
\bstctlcite{IEEEexample:BSTcontrol}

% \bibliographystyle{IEEEtranN}
%\bibliographystyle{IEEEtranS}
%\bibliography{IEEEabrv, main}

% Include the BBL file manually
% Generated by IEEEtranS.bst, version: 1.14 (2015/08/26)

\end{document}